\documentclass[journal,twoside,web]{ieeecolor}
\usepackage{tmi}
\usepackage{colortbl}
\usepackage{amsmath,amssymb,amsfonts}
\usepackage{algorithmic}
\usepackage{graphicx}
\usepackage{textcomp}
\usepackage{placeins}
\usepackage{biblatex} 
\usepackage[linesnumbered,ruled,vlined]{algorithm2e}
\usepackage{tikz}
\usepackage[hidelinks]{hyperref}
\usepackage{xcolor}

\newcommand{\cmark}{\textcolor{green}{\checkmark}}
\newcommand{\xmark}{\textcolor{red}{\texttimes}}

\def\BibTeX{{\rm B\kern-.05em{\sc i\kern-.025em b}\kern-.08em
    T\kern-.1667em\lower.7ex\hbox{E}\kern-.125emX}}

\begin{document}

\title{MedicoSAM: Robust Improvement of SAM for Medical Imaging}

\author{Anwai Archit, Luca Freckmann, Constantin Pape
\thanks{Institute of Computer Science, University of Göttingen, Germany}
}

\maketitle

\begin{abstract}
Medical image segmentation is an important analysis task in clinical practice and research. Deep learning has massively advanced the field, but current approaches are mostly based on models trained for a specific task. Training such models or adapting them to a new condition is costly due to the need for labeled data. The emergence of vision foundation models, especially Segment Anything Model (SAM), offers a path to universal segmentation for medical images, overcoming these issues.
Here, we study how to improve SAM for medical images by comparing different finetuning strategies on a large and diverse dataset. We evaluate the finetuned models on a wide range of interactive and automatic semantic segmentation tasks. 
We find that performance clearly improves given the correct choice of finetuning strategies. This improvement is especially pronounced for interactive segmentation. Semantic segmentation also benefits, but the advantage over traditional segmentation approaches is inconsistent.
Our best model, MedicoSAM, is publicly available. We show that it is compatible with existing tools for data annotation and believe that it will be of great practical value.
\end{abstract}


\section{Introduction} \label{sec:introduction}

Foundation models are large deep neural networks, often based on the transformer architecture \cite{vaswani2017attention}, trained on diverse datasets, either with a self-supervised or supervised objective.
They learn powerful representations that enable different downstream tasks either through in-context learning or finetuning.
They underlay recent advances in language processing \cite{brown2020language} and are also gaining importance in computer vision, thanks to the vision transformer (ViT) \cite{dosovitskiy2020image}.
The first foundation model that has gained wide-spread adoption for image segmentation is the Segment Anything Model (SAM) \cite{kirillov2023segment}. It was trained on a large dataset of natural images with object annotations, using a supervised training objective that mimics interactive annotation. The model supports interactive and automatic segmentation tasks and generalizes to many different imaging modalities. More recently, SAM2 \cite{ravi2024sam} has extended SAM to video data through architectural changes and a large video dataset with objects tracked over time. It supports interactive video segmentation in different modalities.

SAM has been widely studied by the medical imaging community. Initial work has evaluated it for medical segmentation tasks (e.g. \cite{mazurowski2023segment, ji2024segment, he2023computer, Zhang2024, hhli2024}). The model showed impressive performance given that it was predominantly trained on natural images, but could not yet compete with domain specific models, especially for difficult tasks such as spine segmentation in MRI \cite{mazurowski2023segment}, segmentation of small organs in CT \cite{huang2024segment}, and other examples \cite{cheng2023sam,he2023computer}.
Consequently, follow-up work has improved SAM for medical images, either by finetuning it for interactive segmentation \cite{ma2024segment,cheng2023sam,wang2023yanzhou,li2024polyp} or by using it as a pretrained encoder for semantic segmentation \cite{zhang2023customized,wu2023medical,chen2023ma,gu2024build,yue2024surgicalsam,bui2023sam3d}. The model has also been adapted in related domains, for example to improve segmentation in microscopy \cite{archit2023segment} and histopathology \cite{horst2024cellvit, pathosam}. Some work \cite{ma2024segment,cheng2023sam,gu2024build} also tried to build a better \emph{foundation model for medical images}, by finetuning SAM on a large medical dataset and publishing the updated model weights.
Some studies have also investigated SAM2 for medical images \cite{dong2024segment,zhu2024medical,ma2024segment2}. They found mixed results for 2D segmentation, improving over SAM for some modalities but with worse performance for others, and promising results for video segmentation \cite{liu2024surgical}.

\begin{figure*}[ht]
\centerline{\includegraphics[width=2\columnwidth]{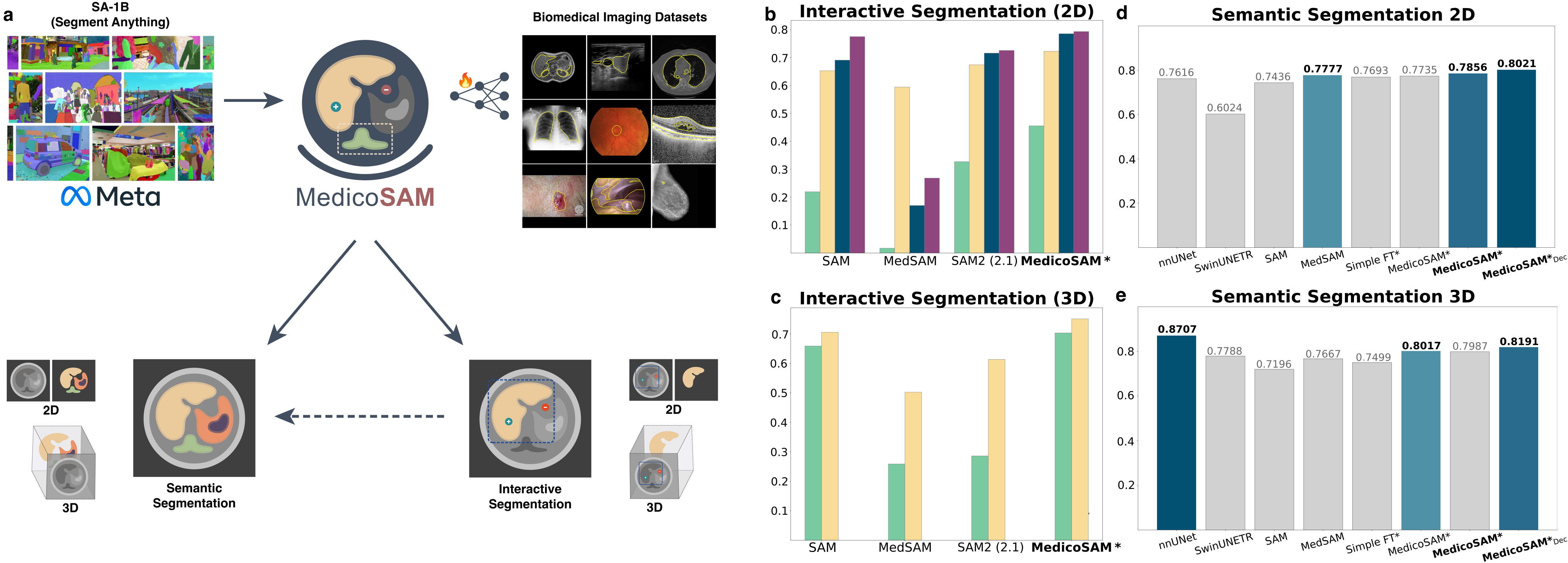}}
\caption{\textbf{a)} Contribution overview: We finetune SAM on a large medical dataset to build MedicoSAM. We evaluate it for interactive and semantic segmentation. The latter requires training on additional annotated data (that could be generated via interactive segmentation), for 2D and 3D data.
\textbf{b)} Results for interactive 2D segmentation, comparing MedicoSAM and other models derived from SAM. We report the average over 16 datasets for segmentation with a point (green) or box (yellow) prompt and segmentation after iterative correction starting from a point (dark green) or box (dark purple). \textbf{c)} Results for interactive 3D segmentation. We report the average over 6 different datasets for segmentation based on a single point or box. \textbf{d, e)}. Results for semantic 2D and 3D segmentation. We report the average over 6 datasets in both cases. The three best methods are highlighted in decreasing shades of blue, darker indicates better results, gray otherwise.}
\label{fig1}
\end{figure*}

However, a comprehensive study that compares different approaches for \emph{improving SAM as a foundation model for medical images} is so far missing. Specifically, no prior work has explored the impact of different finetuning strategies on the different criteria that a foundation model for medical segmentation should fulfill:
\begin{enumerate}
    \item It should improve interactive segmentation. The imaging modalities and segmentation tasks in medicine are very diverse. Hence, a single model that can solve any medical segmentation task automatically is currently not feasible\footnote{There exist some efforts to establish such models for specific modalities, most notably TotalSegmentator \cite{wasserthal2023totalsegmentator,d2024totalsegmentator} for CT and MRI images.}. Better interactive segmentation will enable semi-automatic data annotation, leading to faster annotation times, either for data analysis or model training. 
    \item It should improve the performance for downstream tasks, in particular as a pretrained model for semantic segmentation. This would enable automating segmentation by supervised finetuning, potentially using annotations generated interactively.
    \item It should be compatible with the original SAM library tools for data annotation (e.g. \cite{liu2023samm,archit2023segment}), enabling users to benefit from improved interactive segmentation.
\end{enumerate}
Previous work has only addressed one of these aspects at a time: MedSAM \cite{ma2024segment} and SAMed2D \cite{cheng2023sam} evaluate interactive segmentation (1), Gu et al. \cite{gu2024build} study semantic segmentation (2).
None of them explicitly study compatibility with user-friendly tools (3).

Our work closes this gap by comparing different training strategies on the dataset published by SAMed2d \cite{cheng2023sam} and evaluating their effect on (1-3). Specifically, we evaluate own models trained with different strategies and published SAM-derived models on challenging medical segmentation tasks from four different categories: interactive 2D/3D segmentation and semantic 2D/3D segmentation. Where applicable, we also compare to SAM2 and MedSAM2 \cite{ma2024segment2}.
An overview of our approach is shown in Fig.~\ref{fig1}a and a summary of our results in Fig.~\ref{fig1}b-e.
We find that domain specific finetuning clearly improves interactive segmentation in 2D and 3D, given the right training objective.
For semantic segmentation, pretraining on medical data leads to improved results compared to the original SAM model, with competitive or better performance compared to nnU-Net \cite{isensee2021nnu} in 2D but worse performance in 3D.
Our software and our best model, which we call MedicoSAM, are available at \url{https://github.com/computational-cell-analytics/medico-sam}.

\section{Methods} \label{sec:methods}
We provide a summary of the contributions made by SAM \cite{kirillov2023segment}, focusing on its training objective (Sec.~\ref{sec:methods_sam}). We then describe the finetuning strategies explored in our study (Sec.~\ref{sec:methods_finetune}), including our contribution for pre-training semantic segmentation, our extension of SAM to interactive 3D segmentation (Sec.~\ref{sec:methods_interactive3d}), our methods for 2D and 3D semantic segmentation (Sec.~\ref{sec:methods_semantic}), and our evaluation methodology (Sec.~\ref{sec:methods_eval}).

\subsection{Segment Anything Model} \label{sec:methods_sam}
\begin{figure*}[ht]
\centerline{\includegraphics[width=\textwidth]{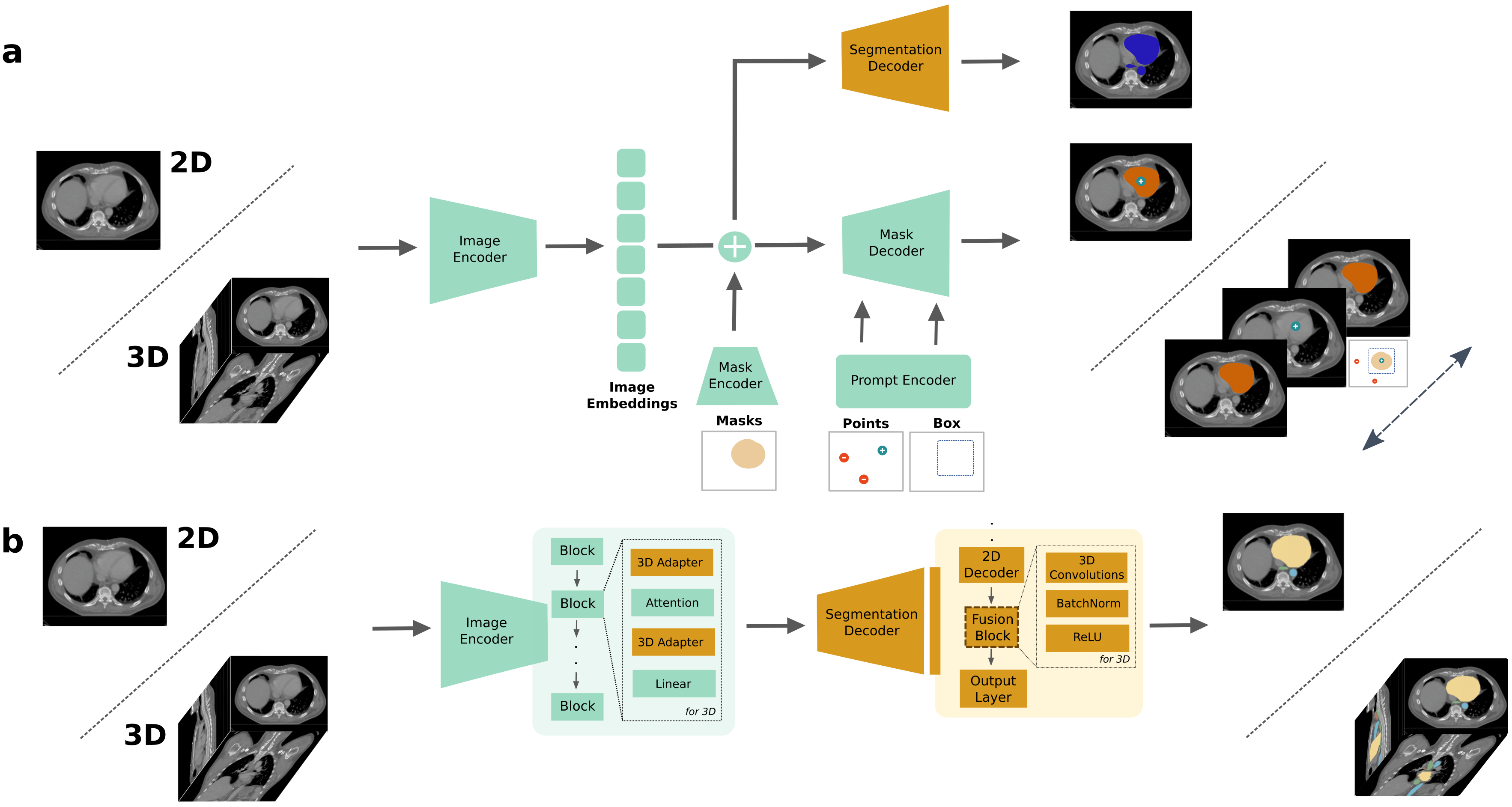}}
\caption{\textbf{a)} The SAM architecture for interactive segmentation consists of image encoder, prompt encoder (split into a part for mask prompts and for point/box prompts), and mask decoder. For 3D interactive segmentation, we propagate prompts across the depth axes. In addition, we add a convolutional decoder for automated segmentation (orange). This decoder is pre-trained with a binary segmentation task (blue masks), jointly with training for interactive segmentation.} \textbf{b)} Adaptation for 2D/3D semantic segmentation. A convolutional decoder predicts the segmentation output (same as "Segmentation Decoder" in a)). For 3D segmentation, additional adapters are added to the image encoder and outputs of the segmentation decoder are computed per slice, stacked, and processed by a 3D convolution. Architecture modifications are highlighted in orange.
\label{fig2}
\end{figure*}

SAM \cite{kirillov2023segment} is a vision foundation model for segmentation tasks.
It consists of the image encoder, a ViT \cite{dosovitskiy2020image}, the prompt encoder and the mask decoder. This architecture enables the model to solve interactive segmentation tasks based on user input, so called prompts. The image encoder processes the image and outputs an image representation. It contains the majority of parameters. SAM provides three different versions with different encoder sizes, ViT-Huge (ViT-h), ViT-Large (ViT-l) and ViT-Base (ViT-b).
The prompt encoder processes the prompts, which can be point coordinates, either a positive point prompt (within the object of interest) or a negative point prompt (outside of the object of interest), a box coordinate, or a low-resolution mask. It outputs a representation of the prompts; point, box and mask prompts can be combined.
The mask decoder processes the outputs of image encoder and prompt encoder to predict a mask of the object of interest and a score that estimates the prediction quality.
It has two heads. One predicts a single mask and score, the other predicts three masks and scores. The second head is for the case of a single point prompt, which can result in ambiguities for part-object segmentation.
Fig.~\ref{fig2}a shows an overview of SAM's architecture with additions made by us marked in orange.

SAM was evaluated for a wide range of segmentation tasks in diverse image modalities, showing remarkable generalization.
These capabilities are mainly due to two factors: its large and diverse training set and sophisticated training objective.
The training dataset, called SA-1B, consists of 11 million images with 1 billion annotated objects.
It was generated by human annotators using SAM for semi-automatic annotation, followed by retraining and further annotation with the updated model, repeated multiple times.
The model was trained on this dataset using a supervised training objective that mimics interactive object annotation and correction:
For a given ground-truth mask, the objective first samples either a point or box prompt and then corrects the model predictions with point prompts in multiple steps. 
In each step it computes $L_{mask}$, the loss between true and predicted mask as well as $L_{iou}$, the loss between the intersection over union (IOU) of true and predicted mask and the predicted score. These losses are accumulated and averaged at the end of a training iteration.
See Alg.~\ref{alg1} for the pseudo-code of a training iteration. Besides this objective, the training proceeds as usual for deep neural networks by updating model weights with a version of stochastic gradient descent over multiple epochs.  
\begin{algorithm}[h]
\DontPrintSemicolon  
\SetArgSty{textnormal}
\begin{small}
\KwIn{Images and target masks, hyperparameters $n_{obj}$, $n_{steps}$, $p_{box}$, $p_{mask}$, $e_{sem}$}
\KwOut{Updated model parameters}
  Sample minibatch of images and target masks\;
  Sample fixed number of object masks $n_{obj}$ per image\;
  Predict embeddings for the images with the encoder\;
  Initialize empty list $L$\ for losses\;
  \For{mask $m$ in minibatch}{
    Initialize empty list for prompts $p$\;
    Sample $u_{box}$ uniformly from $[0, 1]$\;
    \If{$u_{box} < p_{box}$} {
        \tcp{The box can also be distorted}
        Compute bounding box of $m$, add as prompt to $p$\;
    }
    \Else {
        Sample random point from $m$, add as positive point prompt to $p$\;
    }
    Apply prompt encoder to $p$\;
    \If{$p$ contains single point prompt} { 
        Predict masks and IOUs with multi mask head of mask decoder\;
        Select predicted mask $\hat{m}$ and IOU $\hat{i}$ with the highest IOU value\;
    }
    \Else {
        Predict mask $\hat{m}$ and IOU value $\hat{i}$ with single mask head of mask decoder\;
    }
    Compute mask loss $L_{mask}(\hat{m}$, $m$)\;
    Compute IOU $i$ between $\hat{m}$\ and $m$\;
    Compute regression loss $L_{iou}(\hat{i}$, $i$)\;
    Add $L_{mask}$ and $L_{iou}$ to $L$\;
    \For{j = 1 \KwTo $n_{steps}$}{
        Sample positive point from $m \, \& \, ! \hat{m}$, add to $p$\;
        Sample negative point from $! m \, \& \, \hat{m}$, add to $p$\;
        Sample $u_{mask}$ uniformly from $[0, 1]$\;
        Remove mask prompt from $p$ if present\;
        \If{$u_{mask} < p_{mask}$} {
            Add $\hat{m}$ to $p$\;
        }
        Run lines 12-21 with current $p$\;
    }
  }
  \If{$e_{sem}$}{
    Compute binary target $b$ as union of all target masks\;
    Predict binary mask $\hat{b}$ with additional segmentation decoder\;
    Compute $L_{mask}(\hat{b}, b)$ and add it to $L$\;
  }
  Average losses in $L$, perform backprop\;
  Update model parameters via optimizer\;
\end{small}
\caption{
The training objective of SAM \cite{kirillov2023segment}, according to \cite{archit2023segment}. We have added the optional joint pretraining of the segmentation decoder ($e_{sem}=1$).} \label{alg1}
\end{algorithm}

Recently, SAM2 \cite{ravi2024sam} has adapted this approach to interactive video segmentation by adding a memory bank that stores prompts and mask predictions from previous frames. SAM2 was trained on a large annotated video dataset.
While SAM2 is promising in the medical domain to analyze videos or volumetric data, it has so far received fewer attention. 
Some studies \cite{ma2024segment2,dong2024segment,zhu2024medical} have evaluated it and, to our knowledge, only MedSAM2 \cite{ma2024segment2} has provided an improved version of the model.
Here, we include SAM2 and MedSAM2 in the evaluation for interactive segmentation, but do not finetune the model on medical data due to its recency and missing support in tools for medical data annotation.

\subsection{Finetuning Segment Anything} \label{sec:methods_finetune}
We compare, implement, and extend different approaches for improving SAM for medical images by finetuning on a labeled dataset of medical images using a variation of Alg.~\ref{alg1}.
This training algorithm has not been published by SAM \cite{kirillov2023segment}, so each publication has used a custom implementation.
The finetuning strategies also differ in which model parts they update, and whether an adapter-based strategy like LoRA \cite{hu2021lora} is used.

The authors of MedSAM \cite{ma2024segment} assemble a dataset of 1.5 million masks from CT, Endoscopy, MRI, X-Ray and other modalities, based on published data.
They introduce a simple training objective that uses only box prompts, which are derived during training from the mask annotations.
This objective corresponds to $n_{steps}=0, \, p_{box}=1, \, p_{mask}=0$ in Alg.~\ref{alg1}.
They use the ViT-b encoder and update all parameters of the image encoder and mask decoder, freezing the prompt encoder.
The finetuned model is evaluated for interactive segmentation based on box prompts. The model is publicly available.

The authors of SAM-Med2D \cite{cheng2023sam} build the SA-Med2D-20M dataset \cite{ye2023sa}, which consists of 20 million masks in 5 million images. The data covers ten different modalities (CT, Endoscopy, MRI, ultrasound, X-Ray and others) and is collected from published data.
They finetune using with a training objective of $n_{steps}=8, \, p_{box}=0.5, \, p_{mask}=1$. They deviate from Alg.~\ref{alg1} in two ways: they sample either 1, 3, 5 or 9 points per step instead of a single positive and negative one (cf. lines 23-24) and they do not compute gradients for the prompt encoder in the steps corresponding to lines 22-29. They use ViT-b as image encoder and insert a low rank projection layer between attention and feed-forward layer of each transformer block, similar to \cite{wu2023medical}. Only these parameters of the image encoder are updated, others frozen. For prompt encoder and mask decoder all parameters are updated.
In addition, they train the model with a smaller input image size of 256 x 256 pixels instead of 1024 x 1024 pixels used by SAM.
They also study a simpler finetuning strategy called FT-SAM where only the mask decoder is updated during training, image and prompt encoder are frozen.
The finetuned models are evaluated for interactive segmentation and are publicly available. 

The authors of \cite{gu2024build} assemble a dataset comprising 100,000 masks in 300,000 images from CT, MRI, X-Ray and ultrasound.
This dataset also contains unlabeled images used for self-supervised training.
They compare two different finetuning objectives: updating the image encoder in a self-supervised manner using MAE \cite{he2021mae} and using a simple supervised strategy with a box or a single point prompt, corresponding to $n_{steps}=0, \, p_{box}=0.5, \, p_{mask}=0$. They further compare the different model sizes (ViT-b, ViT-l, Vit-h) with and without the use of LoRA \cite{hu2021lora}.
They evaluate these models for semantic segmentation tasks. None of the models are publicly available.

In summary, prior work has studied different finetuning objectives and different model update strategies. To analyze the influence of the objectives we build on the versatile implementation of Alg.~\ref{alg1} provided by $\mu$SAM \cite{archit2023segment}, which was developed for microscopy data.
We extend this implementation to also support simpler schemes (e.g. $n_{steps} = 0$) within the same framework.
We do not study self-supervised training and we finetune all model parts, without the use of adapter layers.
The first choice is due to the fact that self-supervised training would very likely lead to a loss of interactive segmentation performance. The second choice because we want to provide models compatible with the SAM library and tools using it.
Introducing adapter layers would make the model incompatible and thus not practically useful, see also Sec.~\ref{sec:tools}. Consequently, we study three different finetuning strategies:
\begin{itemize}
    \item MedSAM (adapted from \cite{ma2024segment}) uses only a box prompt, corresponding to $n_{steps}=0, \, p_{box}=1, \, p_{mask}=0$ in Alg.~\ref{alg1}.
    \item SimpleFT (adapted from \cite{gu2024build}) uses a single box or a single point prompt, corresponding to $n_{steps}=0, \, p_{box}=0.5, \, p_{mask}=0$ in Alg.~\ref{alg1}.
    \item MedicoSAM uses the full objective with $n_{steps}=8, \, p_{box}=0.5, \, p_{mask}=0.5$\footnote{We have trained two different versions of this model, one with $p_{mask}=0.5$ and one with $p_{mask}=0$, see \label{sec:exp} for details.}.
\end{itemize}
We set $n_{obj}=5$, use the Dice loss for $L_{mask}$ and the L2 loss for $L_{iou}$ in all cases.
We also add an option to jointly pre-train the decoder for semantic segmentation. In this case, the decoder predicts a binary mask and we compute a loss between this prediction and the union of all target masks, corresponding to $e_{sem} = 1$ in Alg.~\ref{alg1}. See Sec.~\ref{sec:methods_semantic} for details on the decoder architecture.
We finetune the models on SA-Med2D-20M \cite{ye2023sa}. We also benchmark the published models MedSAM \cite{ma2024segment}, SAM-Med2D \cite{cheng2023sam}, FT-SAM \cite{cheng2023sam}, SAM \cite{kirillov2023segment}, and where applicable SAM2 \cite{ravi2024sam} and MedSAM2 \cite{ma2024segment2}. We use ViT-b for all models, expect for MedSAM2, where only ViT-Tiny (ViT-t) is available.

\subsection{Interactive 3D Segmentation} \label{sec:methods_interactive3d}
Unlike SAM2, which supports image and video segmentation, SAM can only segment 2D images. Follow-up work has implemented interactive segmentation for videos or volumetric data in medical images, e.g. for CT \cite{lei2023medlsam, maggtraining}.
Here, we use the implementation from \cite{archit2023segment}, which is based on prompt propagation.
Briefly, SAM segments an object in one or multiple slices based on given prompts. Then, the segmentation mask(s) are projected to adjacent slices, prompts are derived from them, and segmentation is run for these prompts. The process is repeated until the object is segmented throughout the whole volume or a stopping criterion based on the IOU between adjacent slices is met.
Multiple options are provided for deriving prompts from projected masks: using a single positive point prompt placed at the mask's center, using multiple point prompts derive from the mask, using the bounding box derived from the mask, using the bounding box and low-resolution version of the mask, and combinations of these options. A user can correct the segmentation by annotating slices with manual prompts and rerunning the segmentation.

\subsection{Semantic Segmentation} \label{sec:methods_semantic}
SAM itself can only be used for interactive segmentation. In \cite{kirillov2023segment} a method for automatic instance segmentation, called automatic mask generation, is proposed. However, it does not support semantic segmentation, which is more relevant for medical images. Conceptually, SAM does not learn explicit semantic knowledge, since it is only trained to distinguish individual objects in an image.
To evaluate different pretrained SAM models for semantic segmentation, we further finetune them for specific tasks using annotated data, based on architectural changes for 2D and 3D segmentation.

For 2D segmentation we add a UNETR-like \cite{unetr} convolutional segmentation decoder. It  processes and upsamples the embeddings predicted by the image encoder, in this case the pretrained ViT-b encoder, to predict a semantic segmentation. Prompt encoder and original mask encoder of SAM are discarded. This model is trained for semantic segmentation with a loss between predictions and semantic labels. Here, we use a combination of Cross Entropy and Dice loss. We also study a variant where the segmentation decoder is pretrained, see Sec.~\ref{sec:methods_finetune} for details. The segmentation decoder is marked in orange in Fig.~\ref{fig2}a. The choice of this architecture for semantic segmentation was motivated by similar approaches for automatic segmentation with SAM in \cite{archit2023segment,sam2unet}.

For 3D segmentation we adopt the implementation of MA-SAM \cite{chen2023ma} to extend the image encoder to volumetric data. This is achieved by flattening the batch and depth dimensions so that the image patches extracted from an input volume can all be processed by the encoder. To make use of depth information, additional adapter layers are introduced. These layers decrease the number of features per token, rearrange tokens into a volumetric representation, apply a 3x1x1 convolution, project the number of features back, and flatten the batch and depth axis.
Each transformer layer is augmented with two of these adapters, one before and one after the attention layer. The parameters of the adapters are randomly initialized.
Unlike MA-SAM, which re-uses SAM's mask decoder, we use our new 2D segmentation decoder. We apply it slice-by-slice to the image embeddings, then stack its outputs and apply a 3D convolution layer to predict a volumetric segmentation.
This design enables to also initialize the pretrained decoder weights. The architecture is shown in Fig.~\ref{fig2}b.

\subsection{Evaluation} \label{sec:methods_eval}
We compare models for interactive and semantic segmentation.
For semantic segmentation (2D and 3D), we follow standard procedures and compare the predicted semantic masks with ground-truth annotations using the Dice coefficient. 

For interactive 2D segmentation we adopt the evaluation procedure of \cite{archit2023segment}.
This approach simulates iterative user-based annotation. It requires object mask annotations. For a given object, a single prompt is sampled, either a point or a box. The object is then iteratively corrected by sampling point prompts from errors in the prediction. In each iteration a positive point prompt is sampled from the region where the prediction is missing (prediction is negative, annotation is positive) and a negative point prompt is sampled from the region where the prediction should not be (prediction is positive, annotation is negative). The Dice coefficient between true mask and prediction is computed for the initial segmentation and each correction iteration.
For interactive 3D segmentation we evaluate the initial segmentation derived from a point prompt (randomly sampled from the object in the central slice) and from a box prompt (also in the central slice).
We do not simulate iterative correction of the masks due to the higher computational demand of 3D segmentation. We run a grid search over the different options for deriving prompts, see Sec.~\ref{sec:methods_interactive3d}, on separate validation data.

\section{Experiments} \label{sec:exp}
We finetune different models based on SAM with ViT-b encoder, initialized with the weights from \cite{kirillov2023segment}: MedSAM, SimpleFT and MedicoSAM, see Sec.~\ref{sec:methods_finetune} for the respective training strategies. The models are trained on the publicly available subset\footnote{The dataset contains a private test set of 920k images and 4 million masks.} of SA-Med2D-20M \cite{ye2023sa}, which contains 3.7 million images and 15.8 million masks.
To compare the impact of $p_{box}$ and $n_{steps}$ (see Alg.~\ref{alg1}), we train a model for each of the three configurations on 60\% of the data (50\% train, 10\% val), using a stratified split over different modalities, setting $p_{mask}=0$, $e_{sem}=0$. We train another version of MedicoSAM on the full dataset (90\% train, 10\% val; same val split as before), with $p_{mask} = 0.5$ and $e_{sem} = 1$. The latter model is included to study the impact of masking and semantic decoder pretraining. The smaller subset for the three initial models is chosen to reduce training cost.
The models are trained on 8 A100 GPUs with 80GB VRAM for 300,000 iterations with a batch size of 7 per GPU, corresponding to 21 epochs for models trained with 60\% data, and 11 epochs for the model trained on the entire data.
We use the AdamW optimizer \cite{loshchilov2017decoupled} with an initial learning rate of $1e-5$ and a scheduler that reduces the learning rate by a factor of 0.9 after each epoch.
We evaluate these four models, the original SAM and SAM2 models, and published derived models for interactive segmentation (2D and 3D), semantic segmentation (2D and 3D), and integration with user-friendly tools.

\begin{figure*}[ht]
\centerline{\includegraphics[width=0.8\textwidth]{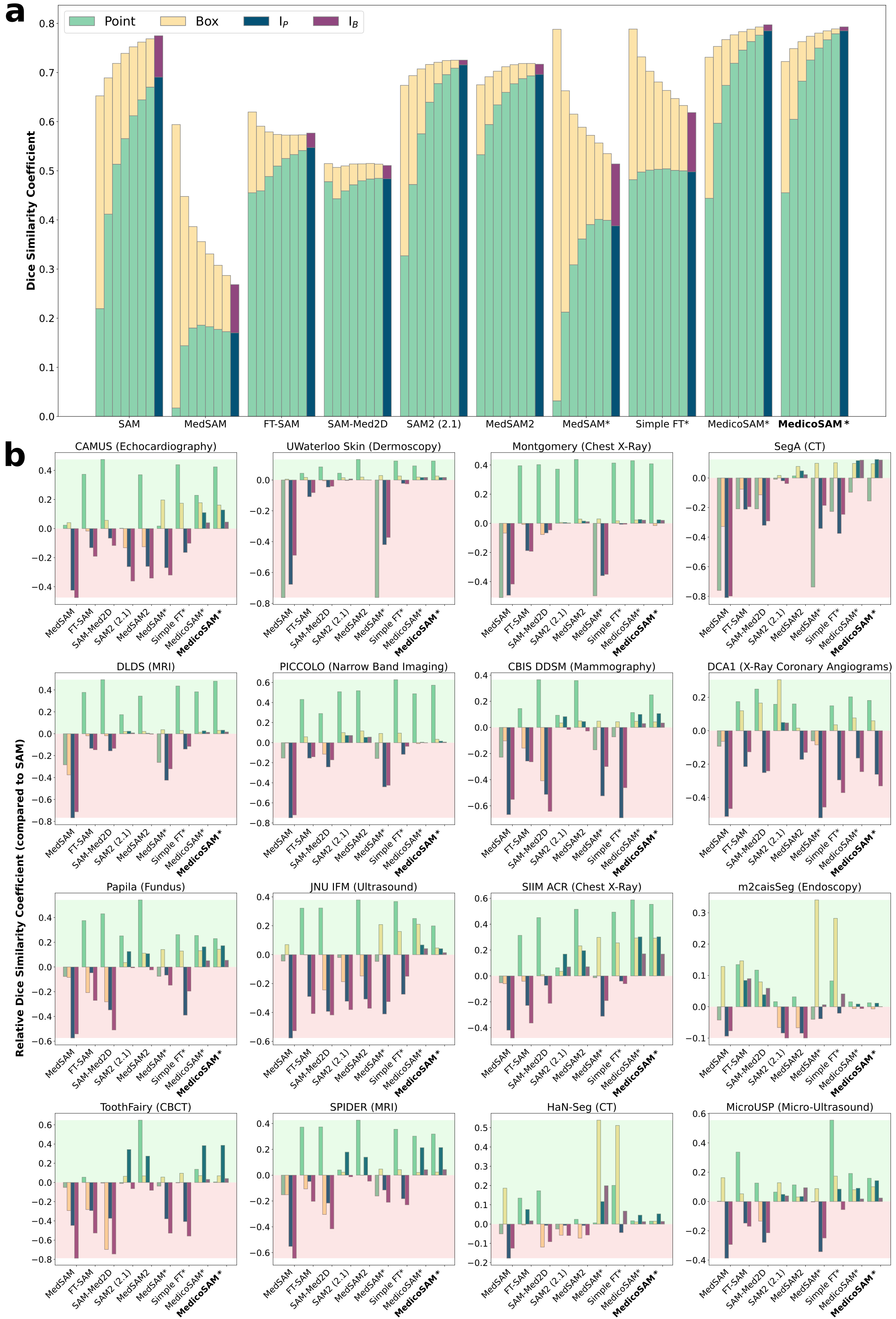}}
\caption{\textbf{a)} Overall results for interactive 2D segmentation. We report the Dice coefficient for simulated interactive segmentation. Each bar corresponds to the result of a correction iteration, starting either from a point (green) or a box (yellow) prompt. The result after correction is highlighted in dark green / dark purple. We compare 10 different models. Models trained by us are marked with a * and the model trained on the entire dataset is marked in bold font. The same model notation is used in all figures. \textbf{b)} Interactive segmentation results for 16 individual datasets. We report the absolute difference of the Dice coefficient compared to the original SAM and report only the results for the initial and final segmentation.}
\label{fig3}
\end{figure*}

\subsection{Interactive Segmentation} \label{sec:exp_interactive}
We evaluate ten different models for interactive 2D segmentation.
We use 16 datasets that are not part of our training dataset to evaluate generalization capabilities. These datasets represent a variety of medical segmentation tasks from CT \cite{pepe2024segmentation,bolelli2023tooth,podobnik2023han}, dermoscopy \cite{glaister2014automatic}, endoscopy \cite{sanchez2020piccolo,maqbool2020m2caiseg}, MRI \cite{macdonald2023duke,van2024lumbar}, ophthalmology \cite{kovalyk2022papila}, ultrasound \cite{leclerc2019deep,lu2022jnu,jiang2024microsegnet}, and X-Ray \cite{jaeger2013automatic,lee2017curated,cervantes2019automatic,siiimacr}. Here, 3D dataset are split into separate images.


The results averaged over all datasets are shown in Fig.~\ref{fig3} a). We report the results for initial prompt-based segmentation and correction iterations, see also Sec.~\ref{sec:methods_eval}, using a point or box as initial prompt. Fig.~\ref{fig3} b) shows the results for individual datasets, where we only report the results for the initial and final segmentation, corresponding to the last correction iteration. We report the difference in the Dice coefficient compared to SAM.
MedicoSAM is the only model that clearly improves upon SAM in all settings. The other finetuned models either only improve for the segmentation with a single prompt (MedSAM*, SimpleFT*, MedSAM2) or lead to an overall worse segmentation. Models trained with $n_{steps} = 0$ yield worse results for an increasing number of prompts.
SAM2 performs slightly worse compared to (original) SAM.

\begin{figure*}[ht]
\centerline{\includegraphics[width=0.95\textwidth]{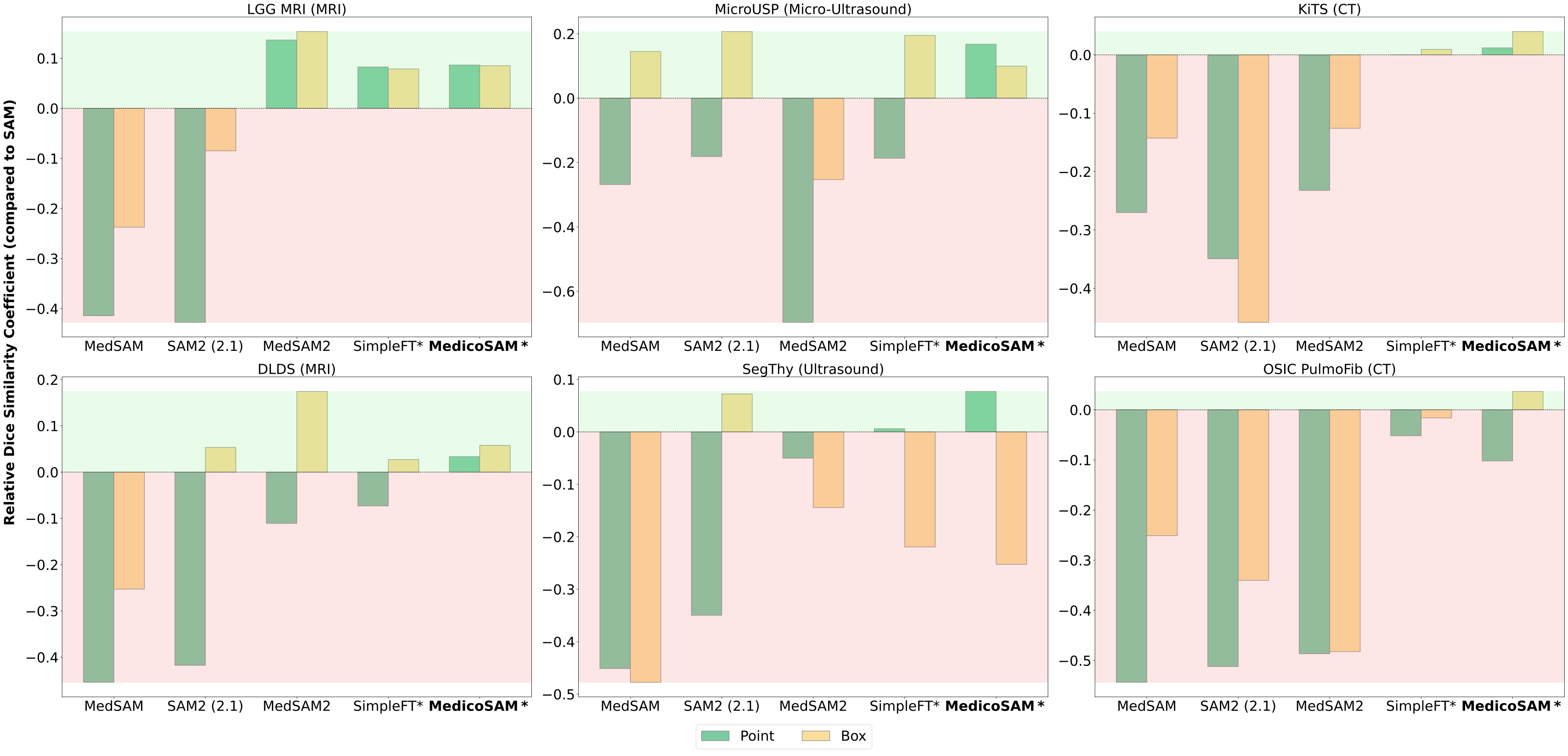}}
\caption{Results for interactive 3D segmentation for 6 different datasets. We report the difference in Dice score compared to SAM for four other models. Segmentations are derived from a single point (green) or box (yellow) prompt placed in the central slice for each object in the respective dataset. We use the implementation of \cite{archit2023segment} for methods using SAM, determining the best method for prompt propagation on a separate validation set, see also Sec.~\ref{sec:methods_interactive3d}.
SAM2 supports 3D segmentation by default.}
\label{fig4}
\end{figure*}

\begin{figure*}[ht]
\centerline{\includegraphics[width=0.8\textwidth]{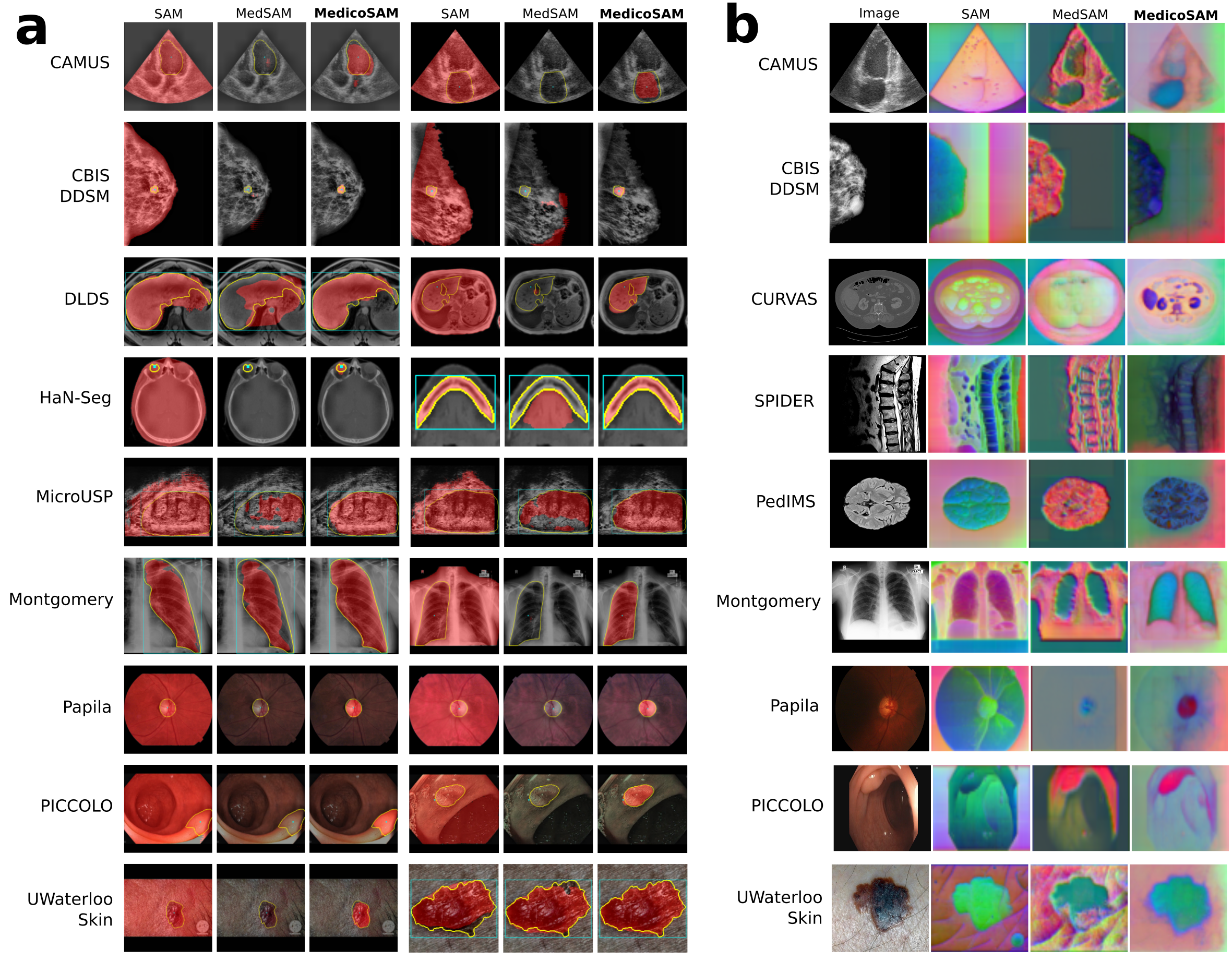}}
\caption{\textbf{a)} Qualitative results for interactive 2D segmentation. We compare interactive segmentation based on a single point or single box prompt (cyan) with SAM, MedSAM and MedicoSAM for nine different datasets. For each image, we show prompts with a large improvement of MedicoSAM over SAM and the corresponding MedSAM result. \textbf{b)} Outputs of the image encoder from the three different models on different datasets, additionally Abdominal CT \cite{curvas} and Brain MRI \cite{pedims} visualized by their three main PCA components. MedSAM and MedicoSAM seem to learn a more discriminative representation with clearer distinction of background.}
\label{appendix_fig1}
\end{figure*}

\begin{figure}[h]
\centering
\includegraphics[width=\columnwidth]{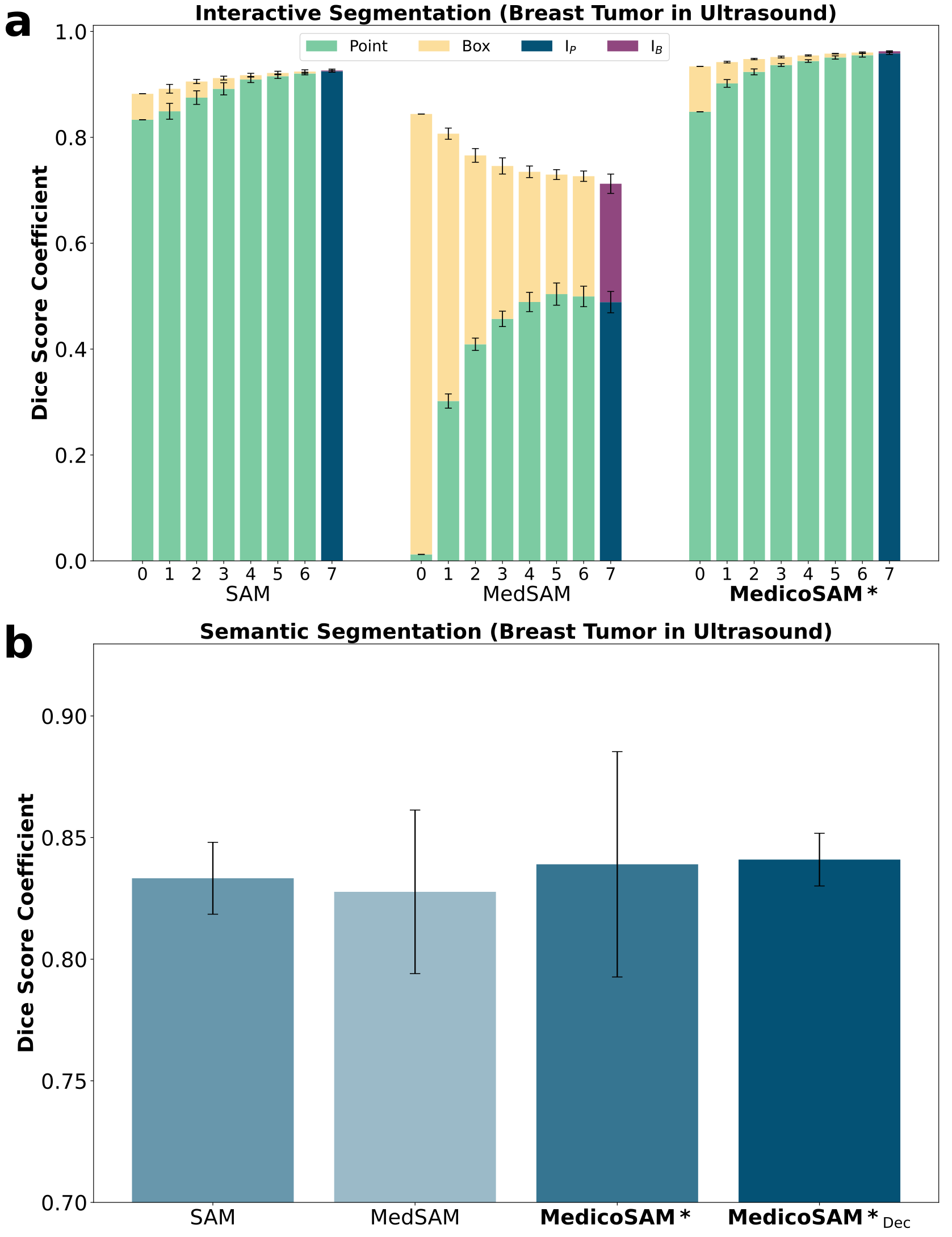}
\caption{\textbf{a)} Statistical analysis of interactive segmentation on the ABUS data for breast tumor segmentation in ultrasound images \cite{abus} with three models. We run each interactive segmentation experiment five times with different random seeds and report standard deviations as error bars. The deviations are small compared to differences in model performance. \textbf{b)} Statistical analysis of semantic segmentation (same data as a)). We run the training for each of the four models five times with different random seeds. The deviations are larger than performance differences, the model with pretrained decoder shows a lower deviation.) The methods are colored in decreasing shades of blue, darker indicates better results.}
\label{fig_statistics}
\end{figure}

\begin{figure}[h]
\centering
\includegraphics[width=\columnwidth]{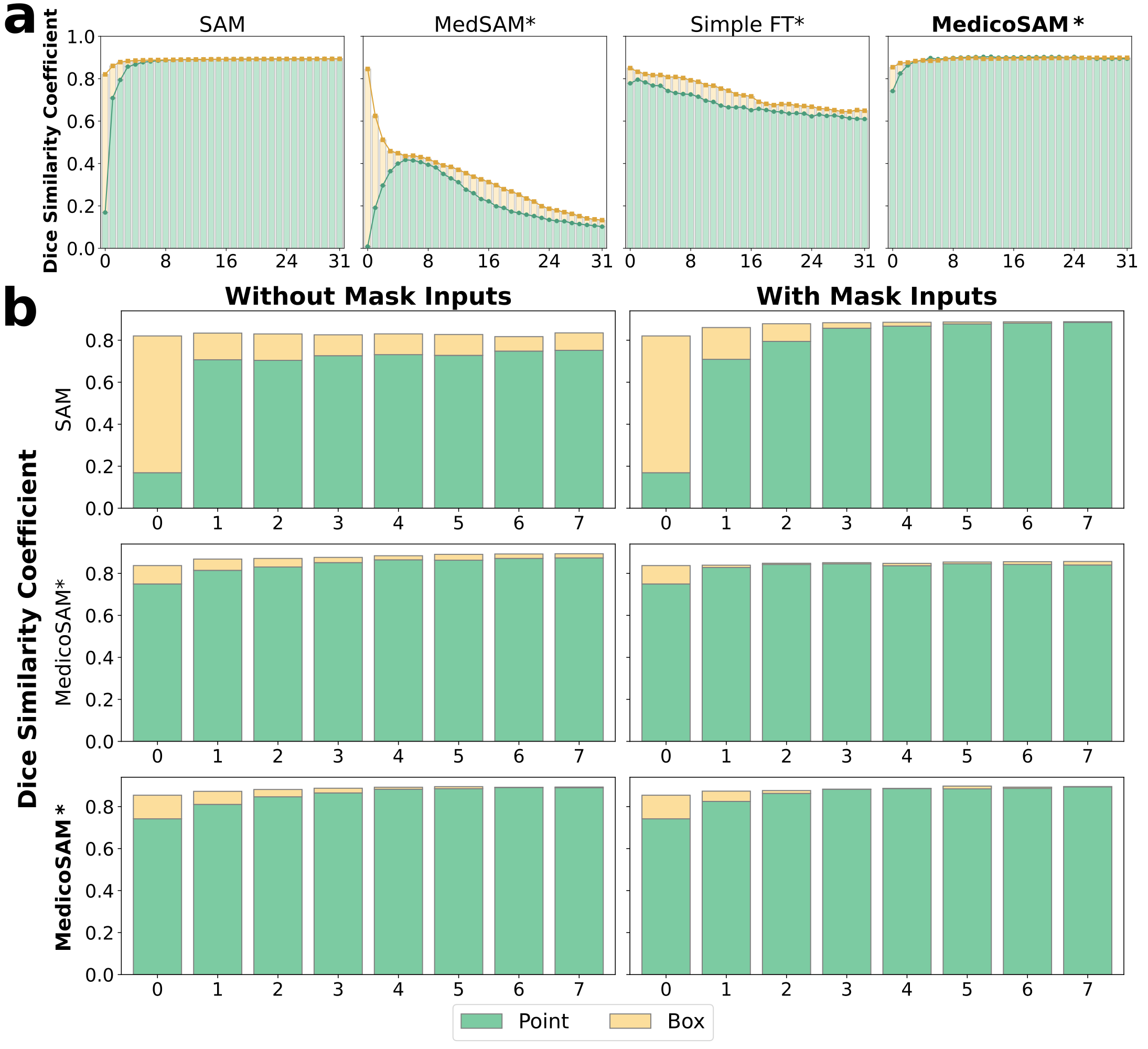}
\caption{\textbf{a)} Ablation study for the influence of $n_{steps}$ in Alg.~\ref{alg1}. We run interactive segmentation with 32 correction iterations, two with $n_{steps} = 0$ (MedSAM, SimpleFT), two with $n_{steps} = 8$. The former show a degrading performance with further prompts, the latter show increasing performance, but plateau after 6-8 iterations. \textbf{b)} Ablation study for the influence of $p_{mask}$, comparing models trained with $p_{mask}$ = 1, 0, 0.5 (SAM, MedicoSAM*, \textbf{MedicoSAM*}) and interactive segmentation without and with use of the previous mask prediction as prompt for the next iteration. SAM and MedicoSAM* perform slightly better in the setting corresponding to their training, \textbf{MedicoSAM*} is robust.}
\label{fig_iterative_prompting}
\end{figure}

A qualitative comparison of results and image embeddings for different models is shown in Fig.~\ref{appendix_fig1}.
We also study the statistical reliability based on five random seeds on a single dataset in Fig.~\ref{fig_statistics}a, where we observe that the standard deviation of results is low compared to differences in the model performance. We perform an additional study to further understand the effect of $n_{steps}$ and $p_{mask}$ in Fig.~\ref{fig_iterative_prompting}.

We evaluate interactive 3D segmentation with a single point or box derived from segmentation annotations for 6 different external datasets from MRI \cite{cancer2015comprehensive,macdonald2023duke}, CT \cite{heller2021state,osic-pulmonary-fibrosis-progression}, and ultrasound \cite{jiang2024microsegnet,kronke2022tracked}. We compare six different models.
For the models based on SAM, we use the method described in Sec.~\ref{sec:methods_interactive3d} and we find the best setting with a grid search on a separate validation set. SAM2 supports interactive 3D segmentation as is (by interpreting the 3D data as a video), and we do not perform a grid search to optimize parameters for inference (same for MedSAM2).
The overall results are shown in Fig.~\ref{fig1} c) (except for SimpleFT) with results for individual datasets in Fig.~\ref{fig4}.
Only MedicoSAM improves consistently.
The fact that SAM2 and MedSAM2 are used as is, without optimizing parameters in a grid search, may disfavor them.

\begin{figure*}[ht]
\centerline{\includegraphics[width=0.9\textwidth]{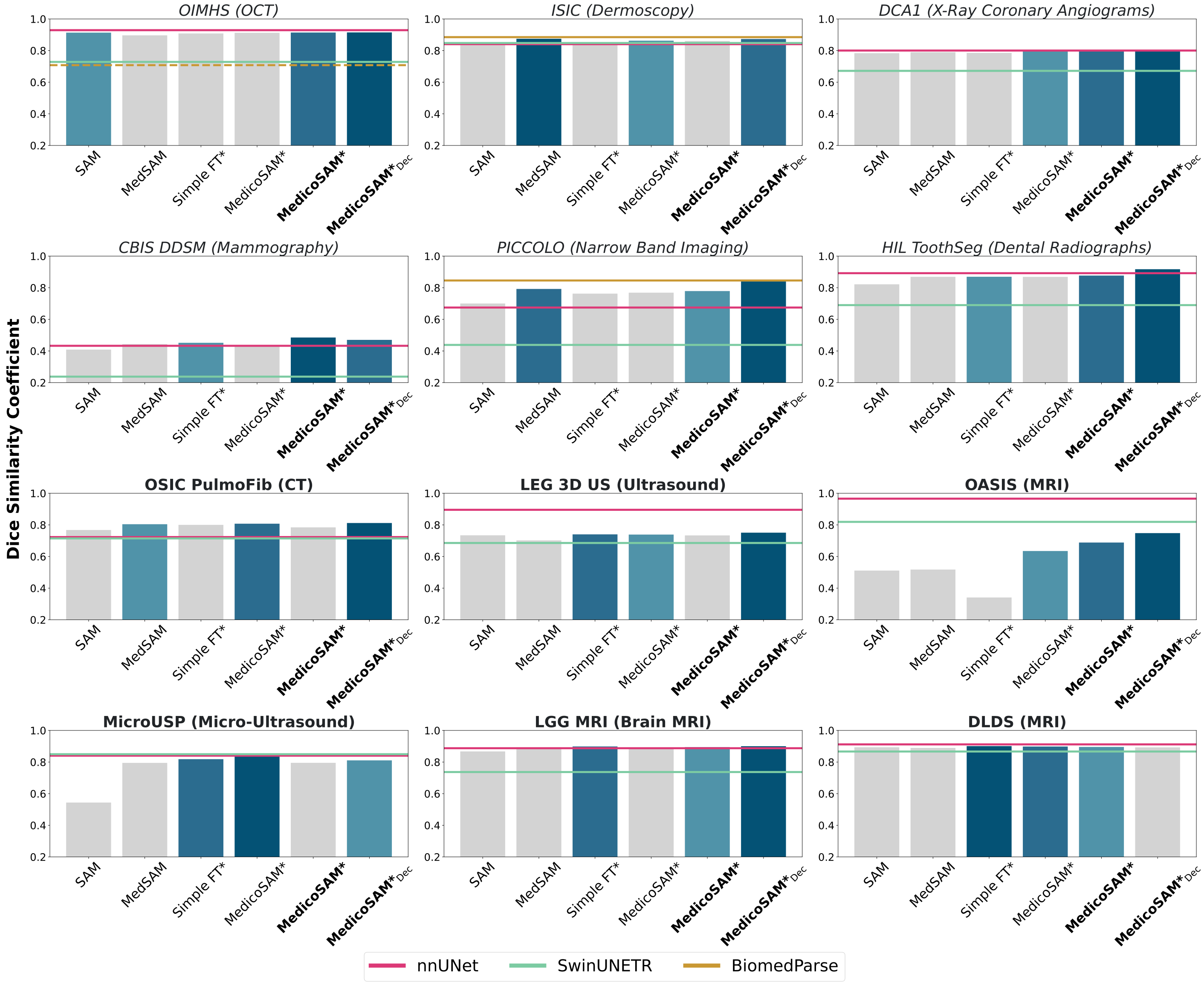}}
\caption{Semantic segmentation results for 2D and 3D segmentation. The names in italic font indicate 2D data and in bold font 3D data. We compare different pretrained models, original SAM, MedSAM, SimpleFT and MedicoSAM in three versions: trained on the reduced training set, trained on the entire training set (bold font) and trained on the entire training set and using decoder initialization (bold font, subscript "Dec"). The three best SAM-derived methods are colored in decreasing shades of blue, darker indicates better results, gray otherwise. Results for reference methods are indicated by horizontal lines. The dotted horizontal line indicates a case where only a subset of classes could be segmented with BiomedParse.}
\label{fig5}
\end{figure*}

\subsection{Semantic Segmentation}
\color{black}
We evaluate different pretrained SAM models for semantic segmentation in 2D and 3D, using the implementations described in Sec.~\ref{sec:methods_semantic}.
For 2D segmentation we use 6 external datasets from dermoscopy \cite{codella2019skin}, mammography \cite{lee2017curated}, narrow band imaging \cite{sanchez2020piccolo}, optical coherence tomography \cite{ye2023oimhs}, panoramic radiographs \cite{roman2021panoramic}, and X-Ray angiography \cite{cervantes2019automatic}. For 3D segmentation we use 6 external datasets from CT \cite{osic-pulmonary-fibrosis-progression}, MRI \cite{macdonald2023duke,marcus2007open,cancer2015comprehensive}, and ultrasound \cite{jiang2024microsegnet,kronke2022tracked}. 
We use separate splits for training and evaluation. 
The results are shown in Fig.~\ref{fig5}. We also report the results for nnU-Net \cite{isensee2021nnu} and Swin UNETR \cite{swinunetr} trained on the same splits, using the default nnU-Net v2 setup and the MONAI \cite{monai} implementation, respectively. We also report the results for BiomedParse \cite{biomedparse}, a foundation model for text-based segmentation trained on a large biomedical imaging dataset. We apply it to the 2D datasets for which we could find a modality prompt matching the training data of BiomedParse.

Here, we see an advantage in using specifically pretrained backbones (MedSAM, SimpleFT, MedicoSAM) compared to the initial SAM model. We also see an advantage in using the pretrained segmentation decoder. These findings are consistent across 2D and 3D datasets.
In 2D, SAM-derived models are competitive or slightly better than nnU-Net and better than Swin UNETR. In 3D, nnU-Net is the best method, in particular due to the poor performance of SAM-based models on two of the datasets. Note that nnU-Net and Swin UNETR are trained from scratch, whereas SAM-derived models are initialized with a pretrained encoder and in one case also a pretrained decoder.
BiomedParse is applied to without  further training and where applicable performs at-par with the best MedicoSAM model.


\subsection{Tool Integration} \label{sec:tools}
\begin{figure*}[!ht]
\centerline{\includegraphics[width=1.0\textwidth]{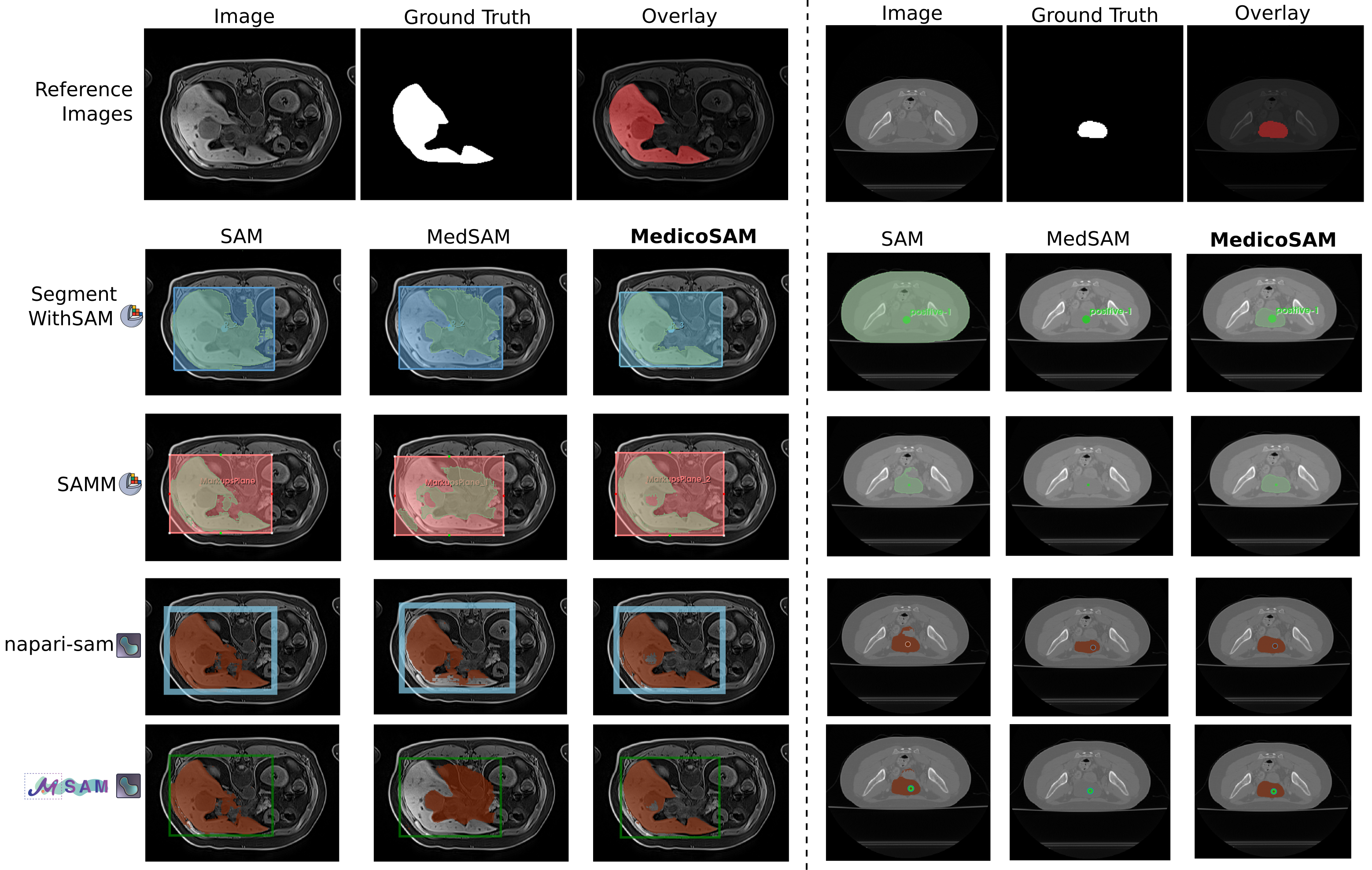}}
\caption{Using SAM, MedSAM and MedicoSAM in four different tools for interactive segmentation. The top row shows the reference image and the object that is annotated. The four rows below show interactive segmentation results with the different tools. Data on the left hand side shows a section from a abdominal MRI scan \cite{ji2022amos}, the right hand side shows a section from a cervical CT scan \cite{landman2015miccai}.}
\label{fig6}
\end{figure*}

An important practical aspect in improving SAM as a foundation model is the integration with tools for data annotation. 
Hence, models should not introduce changes to the architecture that make it incompatible with the SAM library and tools using it. We check this for three models, MedSAM, SAM-Med2D and MedicoSAM, and four different tools: two napari plugins \cite{napari-sam,archit2023segment} and two 3D Slicer extensions \cite{liu2023samm,yildiz2024segmentwithsam}.
Tab.~\ref{tab1} shows the compatibility. We find that SAM-Med2D does not work in any of the tools because it uses adapters and changes the image input size. MedSAM and MedicoSAM work in all of the tools with at most small code changes.
We qualitatively compare the models for data annotation with these tools in Fig.~\ref{fig6}.

\begin{table}[h]
\centering
\begin{tabular}{|l|c|c|c|}
\hline
Tool / Model    & MedSAM & SAM-Med2D & MedicoSAM \\ \hline
SegmentWithSAM  & \cmark & \xmark     & \cmark$^{*}$  \\ \hline
SAMM            & \cmark & \xmark     & \cmark$^{*}$  \\ \hline
napari-sam      & \cmark & \xmark     & \cmark$^{*}$  \\ \hline
$\mu$SAM        & \cmark & \xmark     & \cmark    \\ \hline
\end{tabular}
\caption{Compatibility with user-friendly tools, for three models and four graphical tools supporting SAM for data annotation. The * represents minor code changes necessary to adapt a file path or URL to load different weights.}
\label{tab1}
\end{table}

\section{Conclusion and Discussion}
We have comprehensively studied how to improve SAM \cite{kirillov2023segment} as a foundation model for medical images, by evaluating the impact of different finetuning objectives on interactive and semantic segmentation. We found that interactive segmentation improved clearly, compared to original SAM and SAM2 \cite{ravi2024sam}, but critically dependent on the choice of objective. For semantic segmentation, we found that domain specific pretraining also provides a benefit. However, the segmentation quality is only modestly better than nnU-Net \cite{isensee2021nnu} in 2D and worse in 3D; despite nnU-Net not being pretrained. We also argued that models based on SAM should adhere to the original architecture to enable integration with user-friendly tools. This is especially important due to their improvement in interactive segmentation, which relies on such tools for practical value. We have published our best model, MedicoSAM, and believe that it will be of great practical value for data annotation. Furthermore, we believe that our findings will also prove valuable to adapt recent and future foundation models, e.g. SAM2, to medical imaging.

Our main goal was the comparison of different objectives for finetuning SAM on large medical data. We found that it is crucial to use an objective that trains segmentation with box and mask prompts for multiple training iterations, i.e. $0<p_{box}<1$ and $n_{steps } > 0$ (Alg.~\ref{alg1}). Otherwise we observed a catastrophic forgetting-like effect, where the model did not yield accurate results for interactive segmentation with a point prompt or with multiple prompts, see results for MedSAM and FT-SAM in Fig.~\ref{fig3}. This observation is especially important since MedSAM, the most cited SAM version for medical images, performs worse than the original SAM in many settings due to its choice of objective, which has been independently reproduced, e.g. by MedSAMix \cite{medsamix}. While not explicitly studying other values than $p_{box} = \{0.5, 1\}$, we assume that values smaller than 0.5 would favor the performance with point prompts and vice versa, but would likely not yield to catastrophic forgetting. A value of 0 would likely result in a model that yields poor results in response to box prompts. We found that models trained with a value of $n_{steps} = 8$ were robust to segmentation with a larger number of prompts, i.e. their performance did not decrease with increasing number of prompts as is the case for $n_{steps} = 0$. However, their performance plateaus. See Fig.~\ref{fig_iterative_prompting}a. Training with a larger $n_{steps}$ value could potentially delay the onset of this plateau and increase its height, i.e. lead to better segmentation with more prompts. Finally, we also studied the effect of $p_{mask}$, where we found that models trained with $p_{mask} = 0$ perform better without a mask prompt in interactive segmentation, models trained with $p_{mask} = 1$ perform better with a mask prompt, and models trained with $p_{mask} = 0.5$ are robust. See Fig.~\ref{fig_iterative_prompting}b. While this trend is expected, interestingly, the difference in performance is not very pronounced, i.e. a model trained with $p_{mask} = 1$ does not learn to rely on the presence of a mask prompt and vice versa.

We also studied the effect on semantic segmentation by finetuning adapted architectures for specific segmentation tasks. Here, we found that SAM-derived models performed modestly better than SAM, but that the exact objective used for pretraining them on the large medical dataset did not have a big impact (similar results for MedSAM, MedicoSAM and SimpleFT). A further modest improvement could be gained by using a pretrained segmentation decoder. See results in Fig.~\ref{fig1} d,e and Fig.~\ref{fig4}. Note that the best SAM-derived model performs only modestly better than nnU-Net \cite{isensee2021nnu} in 2D segmentation and worse than in 3D, despite the fact that nnU-Net was not pretrained. This highlights that the advantages of SAM-derived models for semantic segmentation in medical images are not yet clear, at least with our implementation. In this context, the finding of \cite{nnunet-revisited} that transformer-based segmentation architectures generally do not perform better than nnU-Net is also relevant. Further improvements would likely require updates of the adaptation strategy, especially an extension to 3D pretraining for volumetric segmentation.

Another important emergent trend are foundation models that can segment images based on text prompts. They are especially promising for semantic segmentation without the need for further training. We have included a comparison to one such model, BiomedParse \cite{biomedparse} in selected experiments. There are further models, in many cases based on the CLIP \cite{clip} architecture, for medical imaging \cite{zhao-clip, mulmodeseg, clip-msd}. MedCLIP-SAM \cite{medclip-samv2} combines CLIP and SAM for text based segmentation. Hence, a promising future avenue of research is the combination of such a model with our findings, to obtain a foundation model that can address both SAM-style interactive and text-based segmentation.

Other related research, e.g SAM-REF \cite{sam-ref}, has investigated the effect of late and early fusion of image features and prompts. SAM uses a late fusion approach. Early fusion can lead to better segmentation of fine-grained structures, while being less efficient because it couples image and prompt processing. Hybrid approaches can provide a good trade-off. This work could profit from our findings on finetuning objectives, as these are orthogonal to the feature alignment mechanism. Furthermore, recent work on parameter efficient finetuning of SAM for biomedical images has found that strategies like LoRA do not provide better segmentation quality, but can lead to more efficiency \cite{peft-sam}. These findings likely directly apply to semantic segmentation finetuning in our case.

\section*{Acknowledgment}
The work of Anwai Archit was funded by the Deutsche Forschungsgemeinschaft (DFG, German Research Foundation) - PA 4341/2-1. The work of Luca Freckmann was funded by the DFG under Germany’s Excellence Strategy - EXC 2067/1-390729940. 
This work is supported by the Ministry of Science and Culture of Lower Saxony through funds from the program zukunft.niedersachsen of the Volkswagen Foundation for the ’CAIMed – Lower Saxony Center for Artificial Intelligence and Causal Methods in Medicine’ project (grant no. ZN4257).
This work was also supported by the Google Research Scholarship “Vision Foundation Models for Bioimage Segmentation”.
We gratefully acknowledge the computing time granted by the Resource Allocation Board and provided on the supercomputer Emmy at NHR@Göttingen as part of the NHR infrastructure, under the project nim00007.
We would like to thank Sebastian von Haaren for help with improving manuscript figures and data visualizations.

\FloatBarrier
\printbibliography

@article{zhang2023customized,
  title={Customized segment anything model for medical image segmentation},
  author={Zhang, Kaidong and Liu, Dong},
  journal={arXiv},
  year={2023},
}

@article{vaswani2017attention,
  title={Attention is all you need},
  author={Vaswani, Ashish and Shazeer, Noam and Parmar, Niki and Uszkoreit, Jakob and Jones, Llion and Gomez, Aidan N and Kaiser, {\L}ukasz and Polosukhin, Illia},
  journal={NeurIPS},
  year={2017},
}

@inproceedings{
    dosovitskiy2020image,
    title={An Image is Worth 16x16 Words: Transformers for Image Recognition at Scale},
    author={Alexey Dosovitskiy and Lucas Beyer and Alexander Kolesnikov and Dirk Weissenborn and Xiaohua Zhai and Thomas Unterthiner and Mostafa Dehghani and Matthias Minderer and Georg Heigold and Sylvain Gelly and Jakob Uszkoreit and Neil Houlsby},
    booktitle={ICLR},
    year={2021},
}

@inproceedings{kirillov2023segment,
  title={Segment anything},
  author={Kirillov, Alexander and Mintun, Eric and Ravi, Nikhila and Mao, Hanzi and Rolland, Chloe and Gustafson, Laura and Xiao, Tete and Whitehead, Spencer and Berg, Alexander C and Lo, Wan-Yen and others},
  booktitle={ICCV},
  year={2023},
}

@article{mazurowski2023segment,
  title={Segment anything model for medical image analysis: an experimental study},
  author={Mazurowski, Maciej A and Dong, Haoyu and Gu, Hanxue and Yang, Jichen and Konz, Nicholas and Zhang, Yixin},
  journal={Medical Image Analysis},
  year={2023},
}

@article{ji2024segment,
  title={Segment anything is not always perfect: An investigation of sam on different real-world applications},
  author={Ji, Wei and Li, Jingjing and Bi, Qi and Liu, Tingwei and Li, Wenbo and Cheng, Li},
  year={2024},
  journal={Machine Intelligence Research},
}

@article{he2023computer,
  title={Computer-vision benchmark segment-anything model (sam) in medical images: Accuracy in 12 datasets},
  author={He, Sheng and Bao, Rina and Li, Jingpeng and Stout, Jeffrey and Bjornerud, Atle and Grant, P Ellen and Ou, Yangming},
  journal={arXiv},
  year={2023},
}

@article{ma2024segment, 
  title={Segment anything in medical images},
  author={Ma, Jun and He, Yuting and Li, Feifei and Han, Lin and You, Chenyu and Wang, Bo},
  journal={Nature Communications},
  year={2024},
}

@article{cheng2023sam,
  title={Sam-med2d},
  author={Cheng, Junlong and Ye, Jin and Deng, Zhongying and Chen, Jianpin and Li, Tianbin and Wang, Haoyu and Su, Yanzhou and Huang, Ziyan and Chen, Jilong and Jiang, Lei and others},
  journal={arXiv},
  year={2023},
}

@article{wu2023medical,
  title={Medical sam adapter: Adapting segment anything model for medical image segmentation},
  author={Wu, Junde and Ji, Wei and Liu, Yuanpei and Fu, Huazhu and Xu, Min and Xu, Yanwu and Jin, Yueming},
  journal={Medical Image Analysis},
  year={2025},
}

@article{brown2020language,
  title={Language models are few-shot learners},
  author={Brown, Tom and Mann, Benjamin and Ryder, Nick and Subbiah, Melanie and Kaplan, Jared D and Dhariwal, Prafulla and Neelakantan, Arvind and Shyam, Pranav and Sastry, Girish and Askell, Amanda and others},
  journal={NeurIPS},
  year={2020},
}

@article{chen2023ma, 
  title={Ma-sam: Modality-agnostic sam adaptation for 3d medical image segmentation},
  author={Chen, Cheng and Miao, Juzheng and Wu, Dufan and Yan, Zhiling and Kim, Sekeun and Hu, Jiang and Zhong, Aoxiao and Liu, Zhengliang and Sun, Lichao and Li, Xiang and others},
  journal={Medical Image Analysis},
  year={2024},
}

@article{wang2023yanzhou,
  title={SAM-Med3D},
  author={Wang, Haoyu and Guo, Sizheng and Ye, Jin and Deng, Zhongying and Cheng, Junlong and Li, Tianbin and Chen, Jianpin},
  journal={ECCV Workshops},
  year={2024},
}

@article{gu2024build, 
  title={How to build the best medical image segmentation algorithm using foundation models: a comprehensive empirical study with Segment Anything Model},
  author={Gu, Hanxue and Dong, Haoyu and Yang, Jichen and Mazurowski, Maciej A},
  journal={MELBA},
  year={2025},
}

@inproceedings{yue2024surgicalsam,
  title={Surgicalsam: Efficient class promptable surgical instrument segmentation},
  author={Yue, Wenxi and Zhang, Jing and Hu, Kun and Xia, Yong and Luo, Jiebo and Wang, Zhiyong},
  booktitle={AAAI},
  year={2024},
}

@inproceedings{li2024polyp,
  title={Polyp-sam: Transfer sam for polyp segmentation},
  author={Li, Yuheng and Hu, Mingzhe and Yang, Xiaofeng},
  booktitle={Medical Imaging 2024: Computer-Aided Diagnosis},
  year={2024},
  organization={SPIE},
}

@article{archit2023segment,
  title = {Segment Anything for Microscopy},
  journal = {Nature Methods},
  publisher = {Springer Science and Business Media LLC},
  author = {Archit,  Anwai and Freckmann,  Luca and Nair,  Sushmita and Khalid,  Nabeel and Hilt,  Paul and Rajashekar,  Vikas and Freitag,  Marei and Teuber,  Carolin and Spitzner,  Melanie and Tapia Contreras,  Constanza and Buckley,  Genevieve and von Haaren,  Sebastian and Gupta,  Sagnik and Grade,  Marian and Wirth,  Matthias and Schneider,  G\"{u}nter and Dengel,  Andreas and Ahmed,  Sheraz and Pape,  Constantin},
  year = {2025},
}

@article{horst2024cellvit,
  title={Cellvit: Vision transformers for precise cell segmentation and classification},
  author={H{\"o}rst, Fabian and Rempe, Moritz and Heine, Lukas and Seibold, Constantin and Keyl, Julius and Baldini, Giulia and Ugurel, Selma and Siveke, Jens and Gr{\"u}nwald, Barbara and Egger, Jan and others},
  journal={Medical Image Analysis},
  year={2024},
  publisher={Elsevier},
}

@article{wasserthal2023totalsegmentator,
  title={TotalSegmentator: robust segmentation of 104 anatomic structures in CT images},
  author={Wasserthal, Jakob and Breit, Hanns-Christian and Meyer, Manfred T and Pradella, Maurice and Hinck, Daniel and Sauter, Alexander W and Heye, Tobias and Boll, Daniel T and Cyriac, Joshy and Yang, Shan and others},
  journal={Radiology: AI},
  year={2023},
  publisher={Radiological Society of North America},
}

@article{liu2023samm,
  title={Samm (segment any medical model): A 3d slicer integration to sam},
  author={Liu, Yihao and Zhang, Jiaming and She, Zhangcong and Kheradmand, Amir and Armand, Mehran},
  journal={arXiv},
  year={2023},
}

@article{he2021mae,
  title={MAE: Masked autoencoders are scalable vision learners},
  author={He, Kaiming and Chen, Xinlei and Xie, Saining and Li, Yanghao and Doll{\'a}r, Piotr and Girshick, Ross},
  journal={CVPR},
  year={2022},
}

@article{bui2023sam3d,
  title={Sam3d: Segment anything model in volumetric medical images},
  author={Bui, Nhat-Tan and Hoang, Dinh-Hieu and Tran, Minh-Triet and Le, Ngan},
  journal={ISBI},
  year={2024},
}

@article{ye2023sa,
  title={Sa-med2d-20m dataset: Segment anything in 2d medical imaging with 20 million masks},
  author={Ye, Jin and Cheng, Junlong and Chen, Jianpin and Deng, Zhongying and Li, Tianbin and Wang, Haoyu and Su, Yanzhou and Huang, Ziyan and Chen, Jilong and Jiang, Lei and others},
  journal={arXiv},
  year={2023},
}

@article{hu2021lora,
  title={Lora: Low-rank adaptation of large language models},
  author={Hu, Edward J and Shen, Yelong and Wallis, Phillip and Allen-Zhu, Zeyuan and Li, Yuanzhi and Wang, Shean and Wang, Lu and Chen, Weizhu},
  journal={ICLR},
  year={2022},
}

@article{isensee2021nnu,
  title={nnU-Net: a self-configuring method for deep learning-based biomedical image segmentation},
  author={Isensee, Fabian and Jaeger, Paul F and Kohl, Simon AA and Petersen, Jens and Maier-Hein, Klaus H},
  journal={Nature Methods},
  year={2021},
}

@article{pepe2024segmentation, 
  title={Segmentation of the Aorta: Towards the Automatic Segmentation, Modeling, and Meshing of the Aortic Vessel Tree from Multicenter Acquisition},
  author={Pepe, Antonio and Melito, Gian Marco and Egger, Jan},
  journal={MICCAI},
  year={2023},
}

@inproceedings{bolelli2023tooth, 
  title={Tooth fairy: A cone-beam computed tomography segmentation challenge},
  author={Bolelli, Federico and Lumetti, L and Di Bartolomeo, M and Vinayahalingam, S and Anesi, A and van Ginneken, B and Grana, C},
  booktitle={MICCAI},
  year={2023},
}

@article{podobnik2023han,
  title={HaN-Seg: The head and neck organ-at-risk CT and MR segmentation dataset},
  author={Podobnik, Ga{\v{s}}per and Strojan, Primo{\v{z}} and Peterlin, Primo{\v{z}} and Ibragimov, Bulat and Vrtovec, Toma{\v{z}}},
  journal={Medical Physics},
  year={2023},
}

@article {glaister2014automatic, 
  title = {Automatic segmentation of skin lesions from dermatological photographs using a joint probabilistic texture distinctiveness approach},
  author = {J Glaister and A Wong and D A. Clausi},
  journal = {IEEE Transactions on Biomedical Engineering},
  year = {2014},
}

@article{sanchez2020piccolo, 
  title={Piccolo white-light and narrow-band imaging colonoscopic dataset: A performance comparative of models and datasets},
  author={S{\'a}nchez-Peralta, Luisa F and Pagador, J Blas and Pic{\'o}n, Artzai and Calder{\'o}n, {\'A}ngel Jos{\'e} and Polo, Francisco and Andraka, Nagore and Bilbao, Roberto and Glover, Ben and Saratxaga, Cristina L and S{\'a}nchez-Margallo, Francisco M},
  journal={Applied Sciences},
  year={2020},
}

@article{maqbool2020m2caiseg,
  title={m2caiseg: Semantic segmentation of laparoscopic images using convolutional neural networks},
  author={Maqbool, Salman and Riaz, Aqsa and Sajid, Hasan and Hasan, Osman},
  journal={arXiv},
  year={2020},
}

@article{macdonald2023duke,
  title={Duke Liver Dataset: A publicly available liver MRI dataset with liver segmentation masks and series labels},
  author={Macdonald, Jacob A and Zhu, Zhe and Konkel, Brandon and Mazurowski, Maciej A and Wiggins, Walter F and Bashir, Mustafa R},
  journal={Radiology: AI},
  year={2023},
}

@article{van2024lumbar,
  title={Lumbar spine segmentation in MR images: a dataset and a public benchmark},
  author={van der Graaf, Jasper W and van Hooff, Miranda L and Buckens, Constantinus FM and Rutten, Matthieu and van Susante, Job LC and Kroeze, Robert Jan and de Kleuver, Marinus and van Ginneken, Bram and Lessmann, Nikolas},
  journal={Scientific Data},
  year={2024},
}

@article{kovalyk2022papila,
  title={PAPILA: Dataset with fundus images and clinical data of both eyes of the same patient for glaucoma assessment},
  author={Kovalyk, Oleksandr and Morales-S{\'a}nchez, Juan and Verd{\'u}-Monedero, Rafael and Sell{\'e}s-Navarro, Inmaculada and Palaz{\'o}n-Cabanes, Ana and Sancho-G{\'o}mez, Jos{\'e}-Luis},
  journal={Scientific Data},
  year={2022},
}

@article{leclerc2019deep,
  title={Deep learning for segmentation using an open large-scale dataset in 2D echocardiography},
  author={Leclerc, Sarah and Smistad, Erik and Pedrosa, Joao and {\O}stvik, Andreas and Cervenansky, Frederic and Espinosa, Florian and Espeland, Torvald and Berg, Erik Andreas Rye and Jodoin, Pierre-Marc and Grenier, Thomas and others},
  journal={IEEE Transactions in Medical Imaging},
  year={2019},
}

@article{lu2022jnu,
  title={The JNU-IFM dataset for segmenting pubic symphysis-fetal head},
  author={Lu, Yaosheng and Zhou, Mengqiang and Zhi, Dengjiang and Zhou, Minghong and Jiang, Xiaosong and Qiu, Ruiyu and Ou, Zhanhong and Wang, Huijin and Qiu, Di and Zhong, Mei and others},
  journal={Data in Brief},
  year={2022},
}

@article{jiang2024microsegnet,
  title={MicroSegNet: A deep learning approach for prostate segmentation on micro-ultrasound images},
  author={Jiang, Hongxu and Imran, Muhammad and Muralidharan, Preethika and Patel, Anjali and Pensa, Jake and Liang, Muxuan and Benidir, Tarik and Grajo, Joseph R and Joseph, Jason P and Terry, Russell and others},
  journal={Computerized Medical Imaging and Graphics},
  year={2024},
}

@article{jaeger2013automatic,
  title={Automatic tuberculosis screening using chest radiographs},
  author={Jaeger, Stefan and Karargyris, Alexandros and Candemir, Sema and Folio, Les and Siegelman, Jenifer and Callaghan, Fiona and Xue, Zhiyun and Palaniappan, Kannappan and Singh, Rahul K and Antani, Sameer and others},
  journal={IEEE Transactions in Medical Imaging},
  year={2013},
  publisher={IEEE},
}

@article{lee2017curated,
  title={A curated mammography data set for use in computer-aided detection and diagnosis research},
  author={Lee, Rebecca Sawyer and Gimenez, Francisco and Hoogi, Assaf and Miyake, Kanae Kawai and Gorovoy, Mia and Rubin, Daniel L},
  journal={Scientific Data},
  year={2017},
}

@misc{siiimacr,
  author = {Anna Zawacki and Carol Wu and George Shih and Julia Elliott and Mikhail Fomitchev and Mohannad Hussain and ParasLakhani and Phil Culliton and Shunxing Bao},
  title = {SIIM-ACR Pneumothorax Segmentation},
  year = {2019},
  note = {Kaggle},
}

@article{cervantes2019automatic,
  title={Automatic segmentation of coronary arteries in X-ray angiograms using multiscale analysis and artificial neural networks},
  author={Cervantes-Sanchez, Fernando and Cruz-Aceves, Ivan and Hernandez-Aguirre, Arturo and Hernandez-Gonzalez, Martha Alicia and Solorio-Meza, Sergio Eduardo},
  journal={Applied Sciences},
  year={2019},
}

@article{ravi2024sam,
  title={Sam 2: Segment anything in images and videos},
  author={Ravi, Nikhila and Gabeur, Valentin and Hu, Yuan-Ting and Hu, Ronghang and Ryali, Chaitanya and Ma, Tengyu and Khedr, Haitham and R{\"a}dle, Roman and Rolland, Chloe and Gustafson, Laura and others},
  journal={ICLR},
  year={2025},
}

@article{dong2024segment,
  title={Segment anything model 2: an application to 2d and 3d medical images},
  author={Dong, Haoyu and Gu, Hanxue and Chen, Yaqian and Yang, Jichen and Chen, Yuwen and Mazurowski, Maciej A},
  journal={arXiv},
  year={2024},
}

@article{zhu2024medical,
  title={Medical sam 2: Segment medical images as video via segment anything model 2},
  author={Zhu, Jiayuan and Qi, Yunli and Wu, Junde},
  journal={arXiv},
  year={2024},
}

@article{ma2024segment2,
  title={Segment Anything in Medical Images and Videos: Benchmark and Deployment},
  author={Ma, Jun and Kim, Sumin and Li, Feifei and Baharoon, Mohammed and Asakereh, Reza and Lyu, Hongwei and Wang, Bo},
  journal={arXiv},
  year={2024},
}

@article{lei2023medlsam,
  title={Medlsam: Localize and segment anything model for 3d medical images},
  author={Lei, Wenhui and Wei, Xu and Zhang, Xiaofan and Li, Kang and Zhang, Shaoting},
  journal={Medical Image Analysis},
  year={2025},
}

@inproceedings{maggtraining,
  title={Training-free Prompt Placement by Propagation for SAM Predictions in Bone CT Scans},
  author={Magg, Caroline and Verweij, Lukas PE and ter Wee, Maaike A and Buijs, George S and Dobbe, Johannes GG and Streekstra, Geert J and Blankevoort, Leendert and S{\'a}nchez, Clara I},
  booktitle={MIDL},
  year={2024},
}

@article{loshchilov2017decoupled,
  title={Decoupled weight decay regularization},
  author={Ilya Loshchilov and Frank Hutter},
  journal={ICLR},
  year={2019},
}

@article{cancer2015comprehensive,
  title={Comprehensive, integrative genomic analysis of diffuse lower-grade gliomas},
  author={Cancer Genome Atlas Research Network},
  journal={New England Journal of Medicine},
  year={2015},
}

@article{heller2021state,
  title={The state of the art in kidney and kidney tumor segmentation in contrast-enhanced CT imaging: Results of the KiTS19 challenge},
  author={Heller, Nicholas and Isensee, Fabian and Maier-Hein, Klaus H and Hou, Xiaoshuai and Xie, Chunmei and Li, Fengyi and Nan, Yang and Mu, Guangrui and Lin, Zhiyong and Han, Miofei and others},
  journal={Medical Image Analysis},
  year={2021},
}

@misc{osic-pulmonary-fibrosis-progression,
  author = {Ahmed Shahin and Carmela Wegworth and David and Elizabeth Estes and Julia Elliott and Justin Zita and SimonWalsh and Slepetys and Will Cukierski},
  title = {OSIC Pulmonary Fibrosis Progression},
  year = {2020},
  note = {Kaggle},
}

@article{kronke2022tracked,
  title={Tracked 3D ultrasound and deep neural network-based thyroid segmentation reduce interobserver variability in thyroid volumetry},
  author={Kr{\"o}nke, Markus and Eilers, Christine and Dimova, Desislava and K{\"o}hler, Melanie and Buschner, Gabriel and Schweiger, Lilit and Konstantinidou, Lemonia and Makowski, Marcus and Nagarajah, James and Navab, Nassir and others},
  journal={PLOS One},
  year={2022},
}

@article{codella2019skin,
  title={Skin lesion analysis toward melanoma detection 2018: A challenge hosted by the international skin imaging collaboration (isic)},
  author={Codella, Noel and Rotemberg, Veronica and Tschandl, Philipp and Celebi, M Emre and Dusza, Stephen and Gutman, David and Helba, Brian and Kalloo, Aadi and Liopyris, Konstantinos and Marchetti, Michael and others},
  journal={arXiv},
  year={2019},
}

@article{ye2023oimhs,
  title={OIMHS: An Optical Coherence Tomography Image Dataset Based on Macular Hole Manual Segmentation},
  author={Ye, Xin and He, Shucheng and Zhong, Xiaxing and Yu, Jiafeng and Yang, Shangchao and Shen, Yingjiao and Chen, Yiqi and Wang, Yaqi and Huang, Xingru and Shen, Lijun},
  journal={Scientific Data},
  year={2023},
}

@article{roman2021panoramic,
  title={Panoramic dental radiography image enhancement using multiscale mathematical morphology},
  author={Rom{\'a}n, Julio C{\'e}sar Mello and Fretes, Vicente R and Adorno, Carlos G and Silva, Ricardo Gariba and Noguera, Jos{\'e} Luis V{\'a}zquez and Legal-Ayala, Horacio and Mello-Rom{\'a}n, Jorge Daniel and Torres, Ricardo Daniel Escobar and Facon, Jacques},
  journal={Sensors},
  year={2021},
}

@article{marcus2007open,
  title={Open Access Series of Imaging Studies (OASIS): cross-sectional MRI data in young, middle aged, nondemented, and demented older adults},
  author={Marcus, Daniel S and Wang, Tracy H and Parker, Jamie and Csernansky, John G and Morris, John C and Buckner, Randy L},
  journal={Journal of Cognitive Neuroscience},
  year={2007},
}

@misc{napari-sam,
  author = {Gotkowski, Karol},
  title = {napari-sam},
  year = {2020},
  note = {Github},
}

@inproceedings{
  yildiz2024segmentwithsam,
  title={SegmentWith{SAM}: 3D Slicer Extension for Segment Anything Model ({SAM})},
  author={Zafer Yildiz and Hanxue Gu and Jikai Zhang and Jichen Yang and Maciej A Mazurowski},
  booktitle={MIDL},
  year={2024},
}

@article{ji2022amos,
  title={Amos: A large-scale abdominal multi-organ benchmark for versatile medical image segmentation},
  author={Ji, Yuanfeng and Bai, Haotian and Ge, Chongjian and Yang, Jie and Zhu, Ye and Zhang, Ruimao and Li, Zhen and Zhanng, Lingyan and Ma, Wanling and Wan, Xiang and others},
  journal={NeurIPS},
  year={2022},
}

@inproceedings{landman2015miccai,
  title={Miccai multi-atlas labeling beyond the cranial vault--workshop and challenge},
  author={Landman, Bennett and Xu, Zhoubing and Igelsias, J and Styner, Martin and Langerak, Thomas and Klein, Arno},
  booktitle={MICCAI},
  year={2015}
}

@article{huang2024segment,
  title={Segment anything model for medical images?},
  author={Huang, Yuhao and Yang, Xin and Liu, Lian and Zhou, Han and Chang, Ao and Zhou, Xinrui and Chen, Rusi and Yu, Junxuan and Chen, Jiongquan and Chen, Chaoyu and others},
  journal={Medical Image Analysis},
  year={2024},
}

@article{d2024totalsegmentator,
title={TotalSegmentator MRI: Sequence-Independent Segmentation of 59 Anatomical Structures in MR images},
  author={D'Antonoli, Tugba Akinci and Berger, Lucas K and Indrakanti, Ashraya K and Vishwanathan, Nathan and Wei{\ss}, Jakob and Jung, Matthias and Berkarda, Zeynep and Rau, Alexander and Reisert, Marco and K{\"u}stner, Thomas and others},
  journal={arXiv},
  year={2024},
}

@article{liu2024surgical,
  title={Surgical sam 2: Real-time segment anything in surgical video by efficient frame pruning},
  author={Liu, Haofeng and Zhang, Erli and Wu, Junde and Hong, Mingxuan and Jin, Yueming},
  journal={NeurIPS Workshops},
  year={2024}
}

@article{swinunetr,
  author = {Hatamizadeh,  Ali and Nath,  Vishwesh and Tang,  Yucheng and Yang,  Dong and Roth,  Holger and Xu,  Daguang},
  title = {Swin UNETR: Swin Transformers for Semantic Segmentation of Brain Tumors in MRI Images},
  journal = {arXiv},
  year = {2022},
}

@article{Zhang2024,
  title = {Segment anything model for medical image segmentation: Current applications and future directions},
  journal = {Computers in Biology and Medicine},
  author = {Zhang,  Yichi and Shen,  Zhenrong and Jiao,  Rushi},
  year = {2024},
}

@article{hhli2024,
  author = {Lee,  Ho Hin and Gu,  Yu and Zhao,  Theodore and Xu,  Yanbo and Yang,  Jianwei and Usuyama,  Naoto and Wong,  Cliff and Wei,  Mu and Landman,  Bennett A. and Huo,  Yuankai and Santamaria-Pang,  Alberto and Poon,  Hoifung},
  title = {Foundation Models for Biomedical Image Segmentation: A Survey},
  journal = {arXiv},
  year = {2024},
}

@inproceedings{pathosam,
    title={Segment Anything for Histopathology},
    author={Titus Griebel and Anwai Archit and Constantin Pape},
    booktitle={MIDL},
    year={2025},
}

@article{sam2unet,
  author = {Xiong,  Xinyu and Wu,  Zihuang and Tan,  Shuangyi and Li,  Wenxue and Tang,  Feilong and Chen,  Ying and Li,  Siying and Ma,  Jie and Li,  Guanbin},
  title = {SAM2-UNet: Segment Anything 2 Makes Strong Encoder for Natural and Medical Image Segmentation},
  journal = {arXiv},
  year = {2024},
}

@article{unetr,
  title={UNETR: Transformers for 3D Medical Image Segmentation},
  author={Ali Hatamizadeh and Dong Yang and Holger R. Roth and Daguang Xu},
  journal={WACV},
  year={2022},
}

@article{monai,
  title={MONAI: An open-source framework for deep learning in healthcare},
  author={Cardoso, M. Jorge and Li, Wenqi and Brown, Richard and Ma, Nic and Kerfoot, Eric and Wang, Yiheng and Murrey, Benjamin and Myronenko, Andriy and Zhao, Can and Yang, Dong and Nath, Vishwesh and He, Yufan and Xu, Ziyue and Hatamizadeh, Ali and Zhu, Wentao and Liu, Yun and Zheng, Mingxin and Tang, Yucheng and Yang, Isaac and Zephyr, Michael and Hashemian, Behrooz and Alle, Sachidanand and Darestani, Mohammad Zalbagi and Budd, Charlie and Modat, Marc and Vercauteren, Tom and Wang, Guotai and Li, Yiwen and Hu, Yipeng and Fu, Yunguan and Gorman, Benjamin and Johnson, Hans and Genereaux, Brad and Erdal, Barbaros S and Gupta, Vikash and Diaz-Pinto, Andres and Dourson, Andre and Maier-Hein, Lena and Jaeger, Paul F and Baumgartner, Michael and Kalpathy-Cramer, Jayashree and Flores, Mona and Kirby, Justin and Cooper, Lee A D and Roth, Holger R and Xu, Daguang and Bericat, David and Floca, Ralf and Zhou, S. Kevin and Shuaib, Haris and Farahani, Keyvan and Maier-Hein, Klaus H and Aylward, Stephen and Dogra, Prerna and Ourselin, Sebastien and Feng, Andrew},
  journal={arXiv},
  year={2022}
}

@article{biomedparse,
  title = {A foundation model for joint segmentation,  detection and recognition of biomedical objects across nine modalities},
  journal = {Nature Methods},
  author = {Zhao,  Theodore and Gu,  Yu and Yang,  Jianwei and Usuyama,  Naoto and Lee,  Ho Hin and Kiblawi,  Sid and Naumann,  Tristan and Gao,  Jianfeng and Crabtree,  Angela and Abel,  Jacob and Moung-Wen,  Christine and Piening,  Brian and Bifulco,  Carlo and Wei,  Mu and Poon,  Hoifung and Wang,  Sheng},
  year = {2024},
}

@article{medsamix,
  author = {Yang,  Yanwu and Su,  Guinan and Hu,  Jiesi and Sammarco,  Francesco and Geiping,  Jonas and Wolfers,  Thomas},
  title = {MedSAMix: A Training-Free Model Merging Approach for Medical Image Segmentation},
  journal = {arXiv},
  year = {2025},
}

@inbook{nnunet-revisited,
  title = {nnU-Net Revisited: A Call for Rigorous Validation in 3D Medical Image Segmentation},
  booktitle = {MICCAI},
  author = {Isensee,  Fabian and Wald,  Tassilo and Ulrich,  Constantin and Baumgartner,  Michael and Roy,  Saikat and Maier-Hein,  Klaus and J\"{a}ger,  Paul F.},
  year = {2024},
}

@article{clip,
  author = {Radford,  Alec and Kim,  Jong Wook and Hallacy,  Chris and Ramesh,  Aditya and Goh,  Gabriel and Agarwal,  Sandhini and Sastry,  Girish and Askell,  Amanda and Mishkin,  Pamela and Clark,  Jack and Krueger,  Gretchen and Sutskever,  Ilya},
  title = {Learning Transferable Visual Models From Natural Language Supervision},
  journal = {arXiv},
  year = {2021},
}

@article{zhao-clip,
  author = {Zhao,  Ziheng and Zhang,  Yao and Wu,  Chaoyi and Zhang,  Xiaoman and Zhou,  Xiao and Zhang,  Ya and Wang,  Yanfeng and Xie,  Weidi},
  title = {Large-Vocabulary Segmentation for Medical Images with Text Prompts},
  journal = {arXiv},
  year = {2023},
}

@article{mulmodeseg,
  author = {Li,  Chengyin and Zhu,  Hui and Sultan,  Rafi Ibn and Ebadian,  Hassan Bagher and Khanduri,  Prashant and Indrin,  Chetty and Thind,  Kundan and Zhu,  Dongxiao},
  title = {MulModSeg: Enhancing Unpaired Multi-Modal Medical Image Segmentation with Modality-Conditioned Text Embedding and Alternating Training},
  journal = {WACV},
  year = {2025},
}

@inproceedings{clip-msd,
  title = {CLIP-Driven Universal Model for Organ Segmentation and Tumor Detection},
  booktitle = {ICCV},
  author = {Liu,  Jie and Zhang,  Yixiao and Chen,  Jie-Neng and Xiao,  Junfei and Lu,  Yongyi and Landman,  Bennett A. and Yuan,  Yixuan and Yuille,  Alan and Tang,  Yucheng and Zhou,  Zongwei},
  year = {2023},
}

@article{medclip-samv2,
  author = {Koleilat,  Taha and Asgariandehkordi,  Hojat and Rivaz,  Hassan and Xiao,  Yiming},
  title = {MedCLIP-SAMv2: Towards Universal Text-Driven Medical Image Segmentation},
  journal = {arXiv},
  year = {2024},
}

@article{sam-ref,
  author = {Yu,  Chongkai and Liu,  Ting and Li,  Anqi and Qu,  Xiaochao and Wu,  Chengjing and Liu,  Luoqi and Hu,  Xiaolin},
  title = {SAM-REF: Introducing Image-Prompt Synergy during Interaction for Detail Enhancement in the Segment Anything Model},
  journal = {arXiv},
  year = {2024},
}

@inproceedings{
    peft-sam,
    title={Parameter Efficient Fine-Tuning of Segment Anything Model for Biomedical Imaging},
    author={Carolin Teuber and Anwai Archit and Constantin Pape},
    booktitle={MIDL},
    year={2025},
}

@article{abus,
  title = {Automatic semantic segmentation of breast tumors in ultrasound images based on combining fuzzy logic and deep learning—A feasibility study},
  journal = {PLoS ONE},
  author = {Badawy,  Samir M. and Mohamed,  Abd El-Naser A. and Hefnawy,  Alaa A. and Zidan,  Hassan E. and GadAllah,  Mohammed T. and El-Banby,  Ghada M.},
  year = {2021},
}

@article{curvas,
  author = {Riera-Marin,  Meritxell and K,  Sikha O and Rodriguez-Comas,  Julia and May,  Matthias Stefan and Pan,  Zhaohong and Zhou,  Xiang and Liang,  Xiaokun and Erick,  Franciskus Xaverius and Prenner,  Andrea and Hemon,  Cedric and Boussot,  Valentin and Dillenseger,  Jean-Louis and Nunes,  Jean-Claude and Qayyum,  Abdul and Mazher,  Moona and Niederer,  Steven A and Kushibar,  Kaisar and Martin-Isla,  Carlos and Radeva,  Petia and Lekadir,  Karim and Barfoot,  Theodore and Herrera,  Luis C. Garcia Peraza and Glocker,  Ben and Vercauteren,  Tom and Gago,  Lucas and Englemann,  Justin and Kleiss,  Joy-Marie and Aubanell,  Anton and Antolin,  Andreu and Garcia-Lopez,  Javier and Ballester,  Miguel A. Gonzalez and Galdran,  Adrian},
  title = {Calibration and Uncertainty for multiRater Volume Assessment in multiorgan Segmentation (CURVAS) challenge results},
  journal = {arXiv},
  year = {2025},
}

@article{pedims,
  title = {PediMS: A Pediatric Multiple Sclerosis Lesion Segmentation Dataset},
  journal = {Scientific Data},
  author = {Popa,  Maria and Vișa,  Gabriela Adriana and Șofariu,  Ciprian Radu},
  year = {2025},
}

\clearpage



\end{document}